\documentclass[11pt,a4paper,english,table]{article}
\usepackage[T1]{fontenc}
\usepackage{babel}
\usepackage[includeheadfoot,margin=2cm]{geometry}
\usepackage{tikz}
\usepackage{lettrine}
\usepackage{graphicx}
\usepackage{comment}
\usepackage{enumitem}
\usepackage{geometry}
\usepackage{changepage}
\usepackage{float}
\usepackage{lettrine}
\usepackage{graphicx}
\usepackage{subcaption}
\usepackage[export]{adjustbox}
\usepackage{wrapfig}
\usepackage{multirow}
\usepackage{booktabs}
\usepackage{longtable}
\usepackage{algorithm}
\usepackage{algpseudocode}
\usepackage{threeparttable}

\usepackage[stable]{footmisc}

\usetikzlibrary{positioning}
\usetikzlibrary{shapes.geometric, arrows}
\usepackage[section]{placeins}
\usepackage{listings}
\usepackage{breqn}
\usepackage{hyperref}
\usepackage{nth}
\hypersetup{
    colorlinks=true,
    linkcolor=blue,
    filecolor=blue,      
    urlcolor=blue,
    citecolor=blue
}
\usepackage{apacite}
\usepackage{natbib}
\usepackage{authblk}
\author[a]{Menna Hassan} 
\author[a]{Nourhan Sakr}
\author[b]{Arthur Charpentier}
\affil[a]{The American University in Cairo, Cairo, Egypt} 
\affil[b]{Université du Québec à Montréal, Montréal, Canada}
\date{}    
\title{Government Intervention in Catastrophe Insurance Markets:\\ A Reinforcement Learning Approach\footnotemark\footnotetext{Emails: mennahass@aucegypt.edu (MH), n.sakr@columbia.edu (NS),
charpentier.arthur@uqam.ca (AC)}}
\usepackage{setspace}
\providecommand{\keywords}[1]
{
  \small	
  \textbf{{Keywords---}} #1
}
\begin{document}
\maketitle
\onehalfspacing

\begin{abstract}
   This paper designs a sequential repeated game of a micro-founded society with three types of agents: individuals, insurers, and a government. Nascent to economics literature, we use Reinforcement Learning (RL), closely related to multi-armed bandit problems, to learn the welfare impact of a set of proposed policy interventions per \$1 spent on them. The paper rigorously discusses the desirability of the proposed interventions by comparing them against each other on a case-by-case basis. The paper provides a framework for algorithmic policy evaluation using calibrated theoretical models which can assist in feasibility studies.
\end{abstract} 


\ 



\ 

{\bf Acknowledgements:} Arthur Charpentier acknowledges the financial support of the AXA Research Fund through the joint research initiative {\em use and value of unusual data in actuarial science}, as well as NSERC grant 2019-07077.


\newpage

\normalsize
  \section{Introduction}
  \label{intro}
  The first two decades of the \nth{21} century saw an unusual volume of all kinds of catastrophes worldwide. Some of the unforgettable events included the 9/11 terrorist attacks in 2001, Hurricane Katrina in 2005, the Global Financial Crisis in 2008, Japan's nuclear disaster in 2011, and the Covid-19 pandemic in 2019. Today, catastrophe risk is anything but declining; the frequency and the severity of catastrophes are expected to rise in the future with the rising risk of climate change. Despite that catastrophes, whether they are natural or man-made, have become increasingly common, societies still fail to manage their catastrophe risks and are vulnerable to devastating losses in wealth and lives. Catastrophe risk management can be done in a few ways. A major way, which is the focus of this paper, is to invest in catastrophe insurance. However, empirical evidence shows that catastrophe insurance markets fail to provide adequate levels of protection in many parts of the world \citep{kunreuther2013insurance}. For instance,  homeowners in Florida's hurricane-prone areas decided to cut their \textit{subsidized} home insurance as hurricanes became less frequent in the area \citep{kunreuther2013insurance}. Another case was when insurers cut back on their supply of terrorism coverage following al-Qaeda terrorist attack of September 2001 \citep{kunreuther2013insurance}. Finally, and most recently, the Covid-19 pandemic-related business-disruption losses  amounted to \$1 trillion per \textit{month} in the U.S. \citep{panrisk}. In contrast, all private U.S. P\&C insurers had a total of the \$800 billion of combined capital by the end of 2019 \citep{panrisk}. The repetitive failures of insurance markets in providing adequate catastrophe protection has invited debate on how should the government intervene.

  \subsubsection*{Our Contributions}
  In this paper, we aim to contribute to the debate on how and when should the government use its different intervention policies.  We make \textit{four} contributions to existing literature on the topic. \textit{First}, we develop a comprehensive sequential repeated game, outlined in Section \ref{timeline}, that includes three types of agents, individuals, insurers, and government, who strategically interact together to optimize behavior under catastrophe risk. Our model is the first model, to the best of our knowledge, that models agents' behavior under catastrophe risk in a sequential repeated game-like setting. We show that this model is reflective of many empirical stylized facts of catastrophe insurance markets, in Section \ref{nogov}. \textit{Second}, we model complex aspects of agents' behavior that were not previously considered under a single model in literature. For example, we incorporate, in our game, individuals' behavioral biases that were reported in literature to affect catastrophe decision-making. We also consider aspects like individuals' consumption and saving decisions under catastrophe risk. On the insurer side, in addition to profit maximization considerations, we look at insurers' market entry and exit and responses to competitive pressures in the catastrophe insurance markets. We also support heterogeneous individual and insurer behavior in our game. \textit{Third}, we allow our government agent to derive optimal intervention in this game by reinforcement learning (RL), a sequential learning family of machine learning (ML) models. An RL model fits very well in solving for optimal government intervention given the complexity of our model. As described above, our model incorporates many aspects of individual and insurer behavior which are characteristic of catastrophe insurance markets. While the consideration of these aspects helps us comprehensively model catastrophe insurance markets, they increase the non-linear behavior of the model and make it complicated to derive a closed-form analytical solution. Certain RL algorithms, like Q-learning, are not only powerful in solving complex systems but also guarantee convergence to a global optimum. Therefore, we employ Q-learning in this paper to derive the optimal government intervention policy by repeated learning or ``trial and error''  (see Section \ref{qlearning} for fitting Q-learning to our model). \textit{Fourth}, we evaluate the quality of policy interventions using the $MVPF$ approach. While the approach is mainly applied for policy evaluation in empirical studies, we apply the $MVPF$ methodology in our theoretic framework. We explain, in Section \ref{observe}, that the $MVPF$ is a great fit for our purposes as the approach measures not only the direct effects of a policy change but also the long-term behavioral responses related to it. As RL models maximize the discounted sum of future rewards, we used the $MVPF$ as the reward function in our Q-learning model to learn more about the long-run implications of policy changes.
  
  \subsubsection*{Problems in Catastrophe Insurance Markets}
  In tracing the causes behind the failures of catastrophe insurance markets, economists regularly discuss supply-side and demand-side problems. Supply-side problems denote situations where insurer-related problems result in sub-optimal outcomes in catastrophe insurance markets. For example, many private insurers claim that catastrophe risks are uninsurable \citep*{jaffee2006should}. For a risk to be insurable, it needs to have a large number of exposure units, statistically measurable losses, uncorrelated claims, and affordable premium offers \citep*{panrisk}. Catastrophe risks fail to satisfy all of these conditions at varying degrees \citep*{panrisk}. For example, catastrophe risks tend to have few exposure units. This is because many, if not all, members of a society claim losses simultaneously due to a catastrophe experience. Without a large number of exposure units, the law of large numbers collapses \citep*{charpentier2014natural}. This would make catastrophe risk less diversifiable violating the concept of risk-pooling. Another related issue is that catastrophe losses generate correlated claims \citep*{raykov2015catastrophe}. Risks with correlated claims impose huge capital risks on insurers who may not be able to pay the claim liabilities incurred by a large share of their clients at once 
\citep*{cummins2002can}. Furthermore, private  insurers can avoid supplying catastrophe insurance completely if they are unable to measure the frequency and severity of catastrophe losses. Without being able to accurately model expected losses, insurers would not be able to charge fair premiums. Finally, due to all the above reasons, catastrophe insurance, when it is privately supplied, is usually offered at exaggerated premiums well beyond the affordability of most individuals \citep*{kousky2012explaining}. 

In addition to supply-side problems, demand-side problems also contribute to the failures of catastrophe insurance markets. Demand-side problems denote situations where issues on the consumer side result in deficient demand for catastrophe insurance. Consumers' under-purchase of catastrophe insurance was attributed to both rational and irrational causes. A rational cause for the under-purchase of insurance, when it is supplied, can be because it is being offered at unaffordable premium rates \citep*{kousky2012explaining}. Other rational reasons include consumer’s belief that losses due to catastrophe risks will be covered by the government i.e. the Samaritan Dilemma \citep{kunreuther2013insurance}. Demand deficiency can be also explained by bounded rationality: consumers have limited information about their exposure to catastrophe risks and the marginal cost of them getting the right information may exceed their marginal benefit \citep*{kousky2012explaining}. This may cause consumers to act based on behavioral heuristics resulting in an underestimated perception of individual’s exposure to catastrophic risk \citep*{kousky2012explaining}. \cite{meyer2017ostrich} presented six behavioral biases that prevent individuals from taking the rational catastrophe protection decisions. These biases include myopia, amnesia, optimism, inertia, simplification, and herding. We will discuss the role of these biases in detail in Section \ref{popsetup} when we include them in our game.

\subsubsection*{Possible Government Intervention}
The repeated failures of catastrophe insurance markets has invited debate on the government's role in managing society's catastrophe risk. \cite{bruggeman2012insurance} discussed a free market approach where the government would remove such obstacles that prevent the private market from operating smoothly. Under a minimalist intervention approach, the government would focus on relaxing regulations and reforming legal reserve requirements \citep{bruggeman2012insurance}. Empirical evidence on Bangladesh show that reducing credit market imperfections, thus increasing access to micro-credit, significantly affects catastrophe insurance participation \citep{akter2008determinants}. In addition to minimalist approach, the government can also aim for a higher level of intervention and mandate, for example, that catastrophe insurance be bundled with other types of insurance like car and property insurance.This was adopted in France under its ``code des assurances'' through the Act of July 13, 1982. It was also adopted in Belgium though its Acts of May 21, 2003 and September 17, 2005. Through this policy, 90-95\% of the Belgian population are protected against natural-catastrophe risks \citep{bruggeman2012insurance}. Government intervention can also be in the form of public-private partnerships. For example, the government can act as a ``reinsurer of last resort in the same way the U.S. Federal Government acted under terrorism risk \citep{beider2020federal}. In maximum intervention, the government can also directly offer state-provided catastrophe insurance policies itself. An example includes the National Flood Insurance Program (NFIP) which is managed by the U.S. Federal Emergency Management Agency (FEMA). The NFIP, however, is heavily criticised for being severely indebted and failing to adequately cover many of the homeowners in Florida \citep{cummins2006should}. While the federal government aimed to address the indebtedness and the other inefficiencies of the NFIP, these changes would cause the premiums to rise at least five-folds \citep{forbes}.

Government intervention can also be used to target demand-side failures. For example, a government can provide more information to individuals about their real catastrophe risk and amount they need to insure for full coverage \citep{kunreuther2013insurance}. Especially in developing countries, catastrophe awareness can counter demand deficiency for catastrophe insurance. Empirical evidence by \cite{akter2008determinants} on Bangladesh show that household education and occupation significantly affect their likelihood to purchase catastrophe insurance. Another policy suggestion is to use choice architecture and nudges to convince consumers to get insured against catastrophe losses \citep{kunreuther2013insurance}. Given that individuals' behavioral biases tend to clog their rational decision-making, \cite{meyer2017ostrich} recommended a framework called the \textit{behavioral risk audit} for a systematic method of designing demand-side policy in the presence of those biases in individuals. The \textit{behavioral risk audit} is composed of a four-step process: (1) listing the behavioral biases, (2) describing the impacts each bias has on individuals and insurers, (3) analyzing the manifestation of the biases in risk preparedness, and (4) designing remedies for each bias \citep{meyer2017ostrich}.

We summarize the government interventions outlined in literature in Table \ref{tab:pol}.
  \footnotesize
  \begin{longtable}{
|p{10cm}|
p{5cm}|
}

\caption{Government interventions to promote catastrophe preparedness}\\
\toprule
\label{tab:pol}
Intervention&Type\\ \midrule
\endfirsthead
\endhead
Relax regulations and reform legal reserve requirements&supply-side\\
Bundle catastrophe insurance with other types of insurance policies&supply-side\\
Act as a reinsurer of last resort&supply-side\\
Offer state-provided insurance policies&supply-side\\
Provide catastrophe awareness sessions&demand-side\\
Use choice architecture and nudges to influence behavior&demand-side\\
Adopt a behavioral risk audit to design remedies for behavioral biases&demand-side\\
Reduce credit market imperfections&demand-side and supply-side\\
\hline

\end{longtable}
\normalsize

\subsection{Related Work}
The question of organizing catastrophe insurance markets and government intervention was extensively analysed in the literature, many of whom we already cited in the preceding section. Particularly related to modeling catastrophe insurance markets, \cite{kousky2012explaining} rely on optimizing solvency-constrained insurers' supply problem and utility-maximizing individuals' demand problem. After solving for market equilibrium, \cite{kousky2012explaining} find that inadequate catastrophe protection can be due to the large gap between insurer premium prices and individuals' willingness to pay. \cite{charpentier2014natural} developed a game-theoretic framework to model agents in catastrophe insurance markets. They concluded that government intervention should emphasize on capital provision to insurers, regulating premium, and promoting catastrophe risk diversification. In the next sections, we will be adopting and adapting these prior modeling attempts. 

To better link and contextualize the literature, we will reference and describe the related work in those sections where they are most relevant. The paper is organized as follows: Section \ref{envrionment} introduces and sets up our environment, Section \ref{ind} discusses the individual agent, Section \ref{ins} discusses the insurer agent, Section \ref{gov} describes the government agent, Section \ref{nogov}  evaluates the model against some stylized facts of real behavior in free catastrophe insurance markets, Section \ref{qlearning} introduces the Q-learning algorithm used to train the government agent and discusses the learning results, and Section \ref{conclusion} concludes the findings and the contributions of this work. In Section \ref{popsetup} we will talk about the literature on behavioral biases in catastrophe decision making, in Section \ref{risks}, we describe how we incorporate prior results on catastrophe. We also refer to prior work on welfare analysis, particularly the $MVPF$ approach, in Section \ref{gov} and refer to stylised facts on catastrophe insurance markets in Section \ref{nogov}.
 
  \section{The Environment}
  \label{envrionment}
 In reinforcement learning terminology, an environment is defined as the agent's world in which it exists and interacts. It serves as the medium over which our game is played. The agents repeatedly interact with the environment until a terminal state is reached. A single ``play'' of this environment is called an episode. In this section, we will set up our environment, define its agents, and describe the sequence of events in a single episode of the environment.

\subsection{Setup}
\label{setup}
Our environment is composed of three types of agents: (1) a heterogeneous population exposed to catastrophe risk, (2) heterogeneous insurers that offer catastrophe insurance policies to protect the population from catastrophe losses , and (3) a government or social planner that aims to improve the preparedness of the society against future catastrophes.

We call a single play of the environment an \textit{episode} and it runs for a period of $T$ time-steps (see Section \ref{timeline} for episode timeline). For every time-step $t \in T$, an abstract catastrophe, natural or man-made, can occur in the environment with probability $\theta$, which represents the catastrophe risk likelihood. In our environment design we simplified ideas about increasing catastrophe risks, as noted in \cite{cummins2006should} for example, and assumed that the catastrophe risk, $\theta$, under the environment is constant over time. To further simplify our environment design and agent behavior, we assumed that there are no inflationary pressures in the environment and the nominal values of our variables, to be introduced in subsequent sections, reflect the real values over time.
\subsection{Episode Timeline}
\label{timeline}
In a single episode of our environment, individuals, insurers, and the government interact and take decisions under catastrophe risk. The following algorithm summarizes the events that occur in a single episode of our environment:
\footnotesize
\begin{algorithm}
  \caption{Sequence of events in a single episode of the environment}
  \begin{algorithmic}[1]
    \For{\text{each time-step $t \in T$}}
    \State \text{A Bernoulli trial is run with $\theta$ as probability of catastrophe }
    \If {\text{Bernoulli trial was successful}}
    \State \text{Catastrophe occurs in society}
    \State \text{Every individual in the population loses a share of their wealth due to catastrophes} 
     \State \text{Individuals who are catastrophe-insured file claims to their insurance companies}
     \State \text{Insurers pay claims to their customers or exit market if insolvent}
    \EndIf
    \State \text{Every individual updates their catastrophe risk perception based on personal assessment}  
    \State \text{Every individual plans their optimal consumption and saving for time-step $t$ and time-step $t+1$}
    \State \text{Every individual determines their catastrophe insurance demand and attempts to buy, renew,} \par\text{or cancel catastrophe insurance policies accordingly}
    
    \State \text{Each insurer collects premiums from clients who were subscribed with them for a year}
    \State \text{Each insurer updates catastrophe loss model and evaluates financial position, deciding}\par\text{whether to stay or leave the market accordingly}
    \EndFor
  \end{algorithmic}
\end{algorithm}
\normalsize
\FloatBarrier
In Sections \ref{ind}, \ref{ins}, and \ref{gov}, we will explain in detail how each agent takes decisions at each step of the episode. Because the main goal of the paper is to evaluate different forms of the government intervention in a society under catastrophe risk, we first need to model the individual and the insurer agent. After sufficiently describing the individual agent in Section \ref{ind} and the insurer agent in Section \ref{ins}, we will proceed to discuss government intervention, in Section \ref{gov}, in the presence of those agents. The reader can skip to Section \ref{gov} directly if they wish to read about government intervention directly.
  \section{The Individual Agent}
  \label{ind}
  The first type of agent supported by our environment is the individual. In this section, we outline how we modeled our individual agent. The individual has a set of attributes and privately derives its optimal behavior from the optimization results of various microeconomic models.
   \subsection{Population Setup}
\label{popsetup}

The population making up the environment is composed of $n$ heterogeneous individuals indexed by $i$. Every individual $i$ belongs to a social class $c \in \{\text{low}, \text{middle}, \text{upper}\}$. Each individual $i$ earns an annual income $Y_i$ which is dependent on the social class they belong to and it remains constant over time (considering no inflation or income raises as we assumed earlier in Section \ref{setup}). At any time $t$, individual $i$ consumes amount $C_{ti}$ from their annual income $Y_{i}$. The excess of individual $i$'s annual income over consumption at any denotes individuals' savings at time $t$, $S_{ti}$. Individuals start with an initial wealth level, i.e. endowment, $W_{0i}$ at time $0$ which accumulates at an annual constant interest rate $r$. At any time $t$, individual $i$'s wealth, $W_{ti}$, is the accumulated sum of savings plus their initial wealth, $W_{0i}$. This is described in more detail in Section \ref{consav}. 

After a catastrophe experience, we define the loss occurring to an individual to be $\lambda_{Ri}W_{ti}$ where $\lambda_{Ri} \in [0,1]$ represents the true proportion of wealth that an individual $i$ loses due to catastrophes. It is important to highlight that we make the distinction between the \textit{real} wealth loss proportion, $\lambda_{Ri}$, and the proportion that the individual \textit{perceives} to lose from their wealth after a catastrophe, $\lambda_{Pi}$ due to inaccurate estimations, we discuss this distinction in detail in Section \ref{demand}. We assume that $\lambda_{Ri}$ is heterogeneous amongst the population to reflect different exposures to risk which is empirically witnessed in societies exposed to catastrophe risk \citep{meyer2017ostrich}. Additionally, we assume that $\lambda_{Ri}$ can change due to individuals' actions. For example, $\lambda_{Ri}$ can increase if individual $i$ behavior becomes less careful due to being insured i.e. a situation of moral hazard.

As was discussed in Section \ref{intro}, individuals tend to under-prepare for catastrophes due to the existence of cognitive behavioral biases that prevent them from making sound rational decisions \citep{meyer2017ostrich}. Related to the discussion on behavioral heuristics, \cite{meyer2017ostrich} identified six behavioral biases that cause individuals to underprepare for catastrophes. These behavioral biases include myopia, amnesia, optimism, inertia, simplification, and herding \citep*{meyer2017ostrich}. Other studies also indicated the existence of the representative heuristic, also called the availability bias or the availability heuristic, in catastrophe insurance decision making \citep{dumm2020representative}. Myopia, amnesia, and optimism biases serve to reduce individuals' risk perception causing them to underestimate catastrophe risk \citep*{meyer2017ostrich}. These biases cause individuals to underestimate the need to get insurance coverage before witnessing a real catastrophe \citep*{meyer2017ostrich}.  Inertia, simplification, and herding biases distort catastrophe risk management decision-making \citep{dumm2020representative}. Inertia bias causes individuals to maintain the status-quo or default option such that they forgo purchasing or renewing catastrophe insurance \citep*{meyer2017ostrich}. Simplification bias causes individuals to make decisions based on a subset of the available, often incomplete, information. This causes them to make incorrect judgements about the amount of insurance coverage they would need \citep*{meyer2017ostrich}. Herding bias causes individuals to make decisions by mimicking their peers or reference group, who may be making irrational risk management decisions themselves \citep*{meyer2017ostrich}. Finally, the representative heuristic causes individuals to have an exaggerated risk perception, at least twice the initial perception, right after witnessing a catastrophe experience \citep{dumm2020representative}. This exaggerated risk perception declines over time as the memory of the last catastrophe experience fades by the amnesia and optimism biases.

In our environment, we consider the representative heuristic (u), the optimism and amnesia biases (o), the myopia bias (m), the simplification bias (f), the inertia bias (n), and the herding bias (h). We represent those behavioral biases in our environment by giving every individual $i$ in the population a bias parameter $\beta_{ik}$ such that $k\in \{\text{u},\text{o},\text{m}, \text{f},\text{n},\text{h\}}$. Each $\beta_{ik} \in [0,1]$ represents the likelihood with which the individual $i$'s decision making is affected by the bias $k$. As was discussed in \cite{meyer2017ostrich}, we assume that individuals' bias parameters $\beta_{ik}$ are a part of the individuals' cognitive makeup and cannot be changed overtime i.e. constants. We assume that each $\beta_{ik}$ captures the pure effect of bias $k$ and the $\beta_{ik}$s are independent. We make this assumption because dependence structures between behavioral biases were seldom modelled in literature and is out of the scope of this work. Modelling dependencies between behavioral biases can be an idea for future work on this research.

A final attribute we gave the individual agent is its catastrophe risk perception at time $t$, $\alpha_{ti}$. We assume that each individual $i$ has a different risk perception, $\alpha_{ti}$, based on their social class $c_i$ (where we assumed that belonging to a social class implies differences in educational attainment levels). Individual $i$'s risk perception at time $t$, $\alpha_{ti}$, evolves based on the individuals' catastrophe experience and behavioral biases we described above (we will discuss the determinants of risk perception in more detail in Section \ref{risks}). We assume that the $\alpha_{ti}$ is, in most time-steps, divergent from the true catastrophe risk $\theta$ such that it either underestimates it or, less often, overestimates it.

As a summary, the Table \ref{tab:popsetup} describes the population setup with the relevant notation: 
\footnotesize
\begin{longtable}{
p{0.35 cm}
p{9.75 cm} p{6cm}
}
\caption{Population Attributes and Notation}\\
 \toprule 
 \label{tab:popsetup}
&Definition&Variability\\ \midrule
\endfirsthead
\caption[]{Population Setup (Continued)}\\
 \toprule  
&Definition&Variability\\ \midrule
\endhead

$c_i$&Social class of individual $i$&Constant\\
\midrule
$Y_i$&Annual income of individual $i$&Constant\\
\midrule
$C_{ti}$&Amount of annual income that individual $i$ consumes at time $t$&Varies at each time-step $t$ (see Section \ref{consav})\\
\midrule
$S_{ti}$&Excess of individual $i$'s annual income $Y_i$ over consumption $C_{ti}$&varies as $C_{ti}$ varies\\
\midrule
$W_{ti}$&The accumulated wealth of individual $i$ at time $t$&varies as more savings happen to individual's account\\
\midrule
$\lambda_{Ri}$&The real proportion of wealth that individual $i$ loses due to catastrophes&Can increase due to moral hazard or decrease by investing in catastrophe protection technology\\
\midrule
$\lambda_{Pi}$&Perceived proportion of wealth that is lost due to catastrophes by individual $i$&Changes with changes in $\lambda_{Ri}$\\
\midrule
$\beta_{ik}$&The likelihood that individual $i$ is going to make decisions affected by bias $k$&Constant\\
\midrule
$\alpha_{ti}$&The catastrophe risk perception of individual $i$ at time $t$&Varies due to catastrophe experience\\
\hline
\end{longtable}
\normalsize

\subsection{Individual Behavior}
 In this section, we will describe the four sub-models governing individual behavior in the environment: (1) model of catastrophe risk perception $\alpha_{ti}$, (2) model of consumption $C_{ti}$, savings $S_{ti}$, and wealth $W_{ti}$, (3) model of catastrophe insurance demand, and (4) model of insurance catastrophe insurance.
\subsubsection{Modelling Catastrophe Risk Perception}
\label{risks}
It is widely acknowledged in literature that multiple psychological biases affect individuals' risk perception as described in Section \ref{intro}. To our knowledge, incorporation of behavioral biases for the determination of individuals' risk perception was not previously done in literature. 

In models of catastrophe insurance markets in literature, ``risk perception" was defined as a probability $p$ that individuals use to get a weighed average of different social outcomes (\citep{kousky2012explaining}, \citep{charpentier2014natural}). However, these models assume that $p$, and its equivalent, are given parameters and do not elaborate as to how it came about.

In this paper, we model individuals' risk perception under two situations: (1) immediately after a catastrophe experience and (2) under no catastrophe experience or when the memory of the last catastrophe faded. Individuals in the environment start with an initial risk perception $\alpha_0$. This risk perception gets updated at every time-step $t$ (see Section \ref{timeline} for a recap episode timeline).

If an individual $i$ just witnessed a catastrophe, their risk perception $\alpha_{t}$ is updated in the next time-step as follows
\begin{equation}\label{eq:ralpha}
    \alpha_{t+1}=\alpha_{t}(1+\beta_{\text{u}})
\end{equation}
where $\beta_{u} \in [2,3]$ denotes individual $i$'s representative heuristic parameter (see Section \ref{popsetup} for a recap on population setup).

Equation (\ref{eq:ralpha}) causes the risk perception $\alpha_{t}$ to at least double after being exposed to a catastrophe. This is consistent with observations reported in the literature by \cite{kunreuther2013insurance}, \cite{atreya2015drives}, and \cite{dumm2020representative}.

If the individual does not have any catastrophe experience or has forgotten the damage caused by the last catastrophe experience, $\alpha_{t}$ is updated in the next time-step as shown by Equation (\ref{eq:oalpha}):
\begin{equation}\label{eq:oalpha}
    \alpha_{t+1}=\alpha_{t}(1-\beta_{\text{o}})
\end{equation}
where $\beta_{o}$ denotes the combined effect of individual $i$'s optimism and amnesia biases. 

Equation (\ref{eq:oalpha}) reflects empirical observations witnessed amongst real populations under catastrophe risk. They tend to underestimate the true catastrophe risk due to optimism or due to a fading memory of past catastrophe experience as discussed in Section \ref{intro}.

\subsubsection{Modelling consumption $C_{ti}$, savings $S_{ti}$, and wealth $W_{ti}$}
\label{consav}

We model individuals' consumption and savings using an intert-temporal consumption-savings model. To the best of our knowledge, our work is the first attempt to integrate consumption and savings to the decision-making of individuals under catastrophe risks. The reason why we allow individuals' to make decisions about consumption and savings is to model changes in individuals' wealth $W_ti$ which is the attribute that is vulnerable to catastrophe shocks. We use a standard multiperiod intertemporal choice model under uncertainty described in many references including \cite{ramsey1928mathematical}, \cite{samuelson1937note}, and \cite{koopmans1960stationary}.

As described in the setup Section \ref{popsetup}, every individual $i$ receives an annual income $Y_i$ and it remains constant over time.
Individuals derive utility from consumption and are assumed to live the entire throughout any given episode of the environment, i.e. from  $t=0, 1,...,T$. This causes individual's $i$ lifetime utility to be represented as follows:
\begin{equation}\label{eq:lifetimeutil}
    U_i=u(C_{0i})+E(\beta_{m} u(C_{1i}))+....+E(\beta_{m}^T u(C_{Ti}))=u(C_{0i})+\Sigma_{t=1}^T \, E(\beta_{m}^tu(C_{ti}))
\end{equation}
where $\beta_m$ represents the individual's myopia bias, where they generate less utility from consumption expected to occur further ahead in the future.

Each individual $i$ aims to derive optimal consumption at time $t$ such that it maximizes utility $U_i$ from Equation (\ref{eq:lifetimeutil}). This optimization problem is constrained by $T$ budget constraints representing each time period as follows:
\begin{equation}\label{eq:intertemporal}
    C_{0i}+S_{0i}=Y_{0i}+W_{0i}
\end{equation}
    $$C_{1i}+S_{1i}=Y_{1i}+W_{1i}$$
    $$C_{2i}+S_{2i}=Y_{2i}+W_{2i}$$
    \begin{centering}
    .
    
    .
    
    \end{centering}
    $$C_{Ti}=Y_{Ti}+W_{Ti}$$
where $W_{ti}$ for any time $t$ is the accumulated sum of savings of past periods plus initial wealth:
\begin{equation}
    W_{ti}=(1+r)^t(W_{0i}+E(S_{0i}))+(1+r)^{t-1}E(S_{1i})+...+(1+r)E(S_{(t-1)i})
\end{equation}

Equations (\ref{eq:lifetimeutil}) and (\ref{eq:intertemporal}) form a constrained optimization problem which can be solved to yield the following optimization condition:
\begin{equation}\label{eq:Euler}
    u'(C_{ti})=\beta_m(1+r)E(u'(C_{(t+1)i}))
\end{equation}
The full proof can be found in many references and we encourage the interested reader to refer to \cite{ramsey1928mathematical}, and \cite{hansen1983stochastic}. Equation (\ref{eq:Euler}) denotes the expectational Euler equation used in situations of uncertainties. 

In our model, we rely on the work of \cite{ikefuji2015expected} to model individuals' utility as a Pareto family utility function\footnotemark \footnotetext{This named was coined by \cite{ikefuji2015expected} who observed that the utility function can be converted into the cumulative distribution function (cdf) of the Pareto distribution by applying a monotonic linear transformation of $z=W_{ti}+\phi$ on Equation \ref{eq:util}.} where $u'>0$ and $u''<0$. We chose to use this family of utility functions because it was proven, by \cite{ikefuji2015expected}, that Pareto family utility functions are the only type of utility functions that satisfy necessary and sufficient conditions for credible consumption assumptions under catastrophe risks. These assumptions include that consumption, $C_{ti}$, needs to be finite and non-negative at every time-step $t$. Therefore, the Pareto utility function is the most-suitable for our purposes, i.e. modelling consumption and saving under catastrophe risks. 

The Pareto family utility function takes the the form described in Equation (\ref{eq:util}). 
\begin{equation}\label{eq:util}
    U(W_{ti})=1-(1+\frac{W_{ti}}{\phi})^{-k}
\end{equation}
where $\phi$ and $k$ are exogenously determined strictly positive hyperparameters.
We refer the interested reader to the work of \cite{ikefuji2015expected} for more details on how they arrived at their conclusions.
\subsubsection{Deriving Catastrophe Insurance Demand}
\label{demand}

In the environment, we define individual $i$'s demand for catastrophe insurance at time-step $t$ to be the maximum amount they are willing to pay to be \textit{fully covered} against potential catastrophes by a one-year insurance contract.\footnotemark \footnotetext{While we are aware that there is a wide variety of insurance coverage options, we assume, for simplicity, that individuals can only get full insurance coverage against their catastrophe risk. This means, if an individual purchases insurance, the insurance policy will cover the total losses an individual expects due to catastrophes. Partial coverage and other packages are not modelled in this paper.} We will refer to this maximum payable amount for insurance by $P_{ti}^{\text{max}}$.

Individuals in the environment can derive the value for $P_{i}^{\text{max}}$ in two ways, determined by the extent towards which their decision-making is affected by the simplification bias $\beta_f$. If the simplification bias dominates an individual's decision-making, they are more likely to set $P^{\text{max}}_{ti}$ using a subset of the information available to the agent \citep{meyer2017ostrich}. 

To determine whether the individual will simplify their decision-making problem, they run a Bernoulli trial where $\beta_f$ is the probability of success. If the Bernoulli trial is successful, the individual chooses to simplify decision-making and sets their $P_{ti}^{\text{max}}$ at time $t$ as follows:

\begin{equation}\label{eq:irrPmax}
    P_{(t+1)i}^{\text{max}}=(1-\beta_h)P_{ti}^{\text{max}}+\beta_h \bar P_{t}^{\text{max}}
\end{equation}
where $\bar P_{t}^{\text{max}}$ is the average maximum payable premium by all other individuals in the environment who are of the same social class as individual $i$ and $\beta_h$ represents  individual $i$'s herding bias. Through Equation (\ref{eq:irrPmax}) individual $i$ makes an irrational decision about their maximum acceptable premium for catastrophe insurance basing it on their herding bias-weighted average of their maximum premium in the last time-step and the average maximum payable premium by their peers. 

If the Bernoulli trial, on the other hand, is unsuccessful, individual $i$ behaves rationally and conducts a rigorous calculation as what would be their maximum payable premium for catastrophe insurance. We rely on classical expected utility theory described in \cite{morgenstern1953theory} and \cite{maccrimon1979utility}. The expected utility theory was used extensively to model demand for catastrophe insurance in literature like in the work of \cite{kousky2012explaining} and \cite{charpentier2014natural}. We adopt ideas from these models and adapt them to suit the setup of our environment.

Under rational decision-making, individuals consider two states of the world and consider their wealth under each state:
\vspace{2pt}

\begin{center}
\begin{tabular}{|c|c|}
\hline
\textbf{State 1: no catastrophes} &\textbf{State 2: catastrophe happens}\\
 Wealth:  $W_{ti}$ & Wealth: $(1-\lambda_{Pi})W_{ti}$\\
 \hline
\end{tabular}
\end{center}
where $\lambda_Pi$ is the individual's perceived loss due to catastrophes computed as follows:
\begin{equation}
    \lambda_{Pi}=(1-\beta_{o})\lambda_{Ri}
\end{equation}
where $\beta_{o}$ is the individual's optimism and amnesia biases.

Individuals calculate their expected utility, Equation (\ref{eq:EU}), of their wealth from states 1 and 2 of the world using their risk perception at time step $t$, $\alpha_{ti}$,modelled in Section \ref{risks}.
\begin{equation}\label{eq:EU}
    E(U)=(1-\alpha_{ti})\,U(W_{ti})+\alpha_{ti}\,U((1-\lambda_{Pi})W_{ti})
\end{equation}
This is where $U$ in Equation (\ref{eq:EU}) is the \cite{ikefuji2015expected} Pareto Family utility function, Equation (\ref{eq:util}), we introduced towards the end of Section \ref{consav}.


From their expected utility, individuals consider their amount of certain wealth, Equation (\ref{eq:CE}), that would yield them the same level of satisfaction as the expected utility in Equation (\ref{eq:EU}). Individuals also consider and their expected wealth from both states of the world, Equation (\ref{eq:EW}).
\begin{equation}\label{eq:CE}
    \text{Certainty Equivalent}=U^{-1}(E(U))
\end{equation}
\begin{equation}\label{eq:EW}
    E(W)=(1-\alpha_{ti})W_{ti}+\alpha_{ti}(1-\lambda_{Pi})W_{ti}
\end{equation}

It would only make sense for rational individuals to purchase insurance up to the amount that is equal to the difference between the expected wealth, Equation (\ref{eq:EW}), and certainty equivalent, Equation (\ref{eq:CE}), assuming that this quantity would be non-negative as individuals are assumed to be risk-averse. If individuals pay more in insurance than the difference between expected wealth and certainty equivalent, they are getting less value from insurance as opposed to experiencing losses from a real catastrophe. Therefore, we set each individual $i$'s maximum payable premium for catastrophe insurance at time-step $t$ under rational decision-making as follows:
\begin{equation}\label{eq:Pmax}
    P^{\text{max}}_{ti}=E(W)-\text{Certainty Equivalent}
\end{equation}
which represents individual's $i$ demand, or willingness to pay, for catastrophe insurance.

\subsubsection{Purchasing Insurance}
\label{buy}
After determining risk perception $\alpha_{ti}$, wealth $W_{ti}$, and demand for insurance $P^{\text{max}}_{ti}$ at time-step $t$, individual $i$ proceeds to purchase catastrophe insurance as long as there is a suitable offer available with in the catastrophe insurance markets.

Each individual $i$ surveys each insurer $j$ for an insurance policy with an annual premium payment that is at most equal to $P^{\text{max}}_{ti}$. If a suitable offer exists, individual $i$ purchases insurance from the suitable insurer who is willing to provide insurance at a premium that is acceptable by them. When the individual successfully purchases an insurance policy, they get an insurance contract with a term of one time-step. The contract is composed of: (1) the agreed premium rate, and (2) the amount of coverage the insurer promises to pay in the event of a catastrophe. 
Therefore, while an individual $i$ is covered by a contract, they are required to pay an annual premium according to Equation (\ref{eq:premium}) in Section \ref{supply}.

Insurance contracts do not automatically renew and if individuals are still interested in coverage at time $t+1$ for time $t+2$, they need to re-purchase insurance policies under new terms considering the updates in $W_{t+1}$, $\alpha_{t+1}$, and $P^{\text{max}}_t$.

While it might be obvious that individuals' decision to renew their insurance coverage at time $t+1$ depends on whether they are able to find an offer that fully covers their wealth at an acceptable premium, we consider situations where individuals fail to renew their coverage simply because they procrastinate to do so and just prefer to "do nothing". The former irrational decision-making is due to individuals' inertia bias, $\beta_n$, that was cited as a possible behavioral explanation as why individuals tend to underinsure \citep{meyer2017ostrich}. When individuals fail to purchase insurance while suitable contracts exist in the environment, it is because of a successful Bernoulli trial that the individual runs before  purchasing insurance with $\beta_n$ as the probability of success.
  \section{The Insurer Agent}
  \label{ins}
The second type of agent in our environment is the insurer. In this section, we will outline how we modeled our insurer agent. Like the individual agent, the insurer agent has a set of attributes and privately determines optimal behavior through underlying models from microeconomic theory.
\subsection{Insurers' Setup}
\label{inssetup}
The environment has $m$ heterogeneous catastrophe insurers, indexed by $j$, who supply insurance policies to protect individuals' wealth from expected losses due to catastrophes. We assume that each insurer $j$ starts with a capital $\kappa_{0j}$. Insurers' capital make up a big part of insurers' assets which they use to supply catastrophe insurance policies (this is described in more detail in Section \ref{supply}). Insurers' capital accumulates at a constant interest rate $r$ per year and the capital gets updated every year as the insurer makes profits or losses. We assume that each insurer $j$ uses proportion $\gamma_j \in [0,1]$ of their assets to supply insurance policies. This assumption was made to reflect heterogeneity in the insurers' internal policies and preferences of the board of directors who may not prefer to use all the assets of the company to supply catastrophe insurance policies. Insurers' $\gamma_j$ is updated after catastrophe exposure as is described in Section \ref{lossmodel}. 

We assume that every insurer $j$ has an exit parameter $\epsilon_j \in [0,1]$ which represents the likelihood the insurer would exit the market after experiencing adverse market conditions. If $\epsilon_j$ reaches a value of 1, insurer $j$ exits the market. We also give every insurer a premium loading rate $l_j$ which is used by insurers to load the risk premium rate for profit maximization. We allow insurers' $l_j$ to vary over time to reflect competition with other insurers in the market. Additionally, each insurer $j$ is given a heterogeneous solvency percentile $\rho_j \in [0,1]$ which is used to calculate the reserves required per policy supplied. Finally, to reflect discussions about insurers' irrational behavior in literature, for example in \cite{kunreuther2013insurance} and \cite{smetters2008financing}, we give each insurer $j$ a bias parameter $\beta'_j \in [0,1]$ which dictates the likelihood with which they are likely to restrict supply and increase premium rates right after catastrophes. Like the individual agent, we assumed that the insurers' bias parameter is a part of its  cognitive makeup and therefore remains constant over time. 

Table \ref{tab:inssetup} summarizes the insurers' setup with the relevant notation:
\footnotesize
\begin{longtable}{
p{0.35 cm}
p{9.75 cm} p{6cm}
}
\caption{Insurers' Attributes and Notation}\\
 \toprule 
 \label{tab:inssetup}
&Definition&Variability\\ \midrule
\endfirsthead
\caption[]{Insurers' Attributes and Notation (Continued)}\\
 \toprule  
&Definition&Status\\ \midrule
\endhead

$\kappa_{tj}$&Capital Available with insurer $j$ at time $t$&Accumulates every year at rate $r$ and changes as insurer makes profits/losses\\
\midrule
$\gamma_j$&Proportion of assets insurer $j$ can use to supply catastrophe insurance policies&Changes after being exposed to catastrophe experience\\
\midrule
$\epsilon_j$&The likelihood with which an insurer $j$ would exit the catastrophe insurance market&Changes with exposure to adverse market conditions\\
\midrule
$l_j$&Insurer $j$ premium loading that is used to load the risk premium rate of insurance policies&Changes due to competitive pressures in the market\\
\midrule
$\rho_j$&The percentile of the distribution of expected losses insurers are required to keep in reserves per policy for solvency&Changes due to catastrophe experience or government regulation\\
\midrule
$\beta'_j$&The bias parameter which represents the likelihood with which insurers would irrationally behave after a catastrophe&Constant Over Time\\
\hline
\end{longtable}
\normalsize
\subsection{Insurer Behavior}
In this section, we will describe three sub-models governing insurers' behavior: (1) choosing a catastrophe loss model, (2) deriving catastrophe insurance supply, and (3) determining market entry and exit.
\subsubsection{Choosing a Catastrophe Loss Model}
\label{lossmodel}
In real-life, catastrophe insurers often hire a professional risk-modelling firm that would provide them with a loss model for the risk they are intending to insure \citep{ericson2004catastrophe}, \citep{grossi2005catastrophe}, and \citep{heinrich2021simulation}. The loss model gives insurers guidance on premium rates they should charge and the amount they should keep in reserves. In our environment, we allow each active insurer $j$ to choose a loss model at time $0$ from a range of heterogeneous catastrophe risk modellers. We assume that the chosen loss model is composed of two items:
\begin{enumerate}
    \item Catastrophe loss probability $p_j$ i.e. risk premium rate
    \item Reserve required per insurance policy supplied
\end{enumerate}
To add realism to our model, we assume that each risk modelling firm does not accurately estimate the actual catastrophe probability $\theta$ (see Section \ref{setup} for a recap on environment setup). Risk modelling firms either overestimate or underestimate the true catastrophe loss probability. Therefore, $p_j\neq\theta$ in the majority of the episodes in our environment.

To recommend a reserve amount per policy supplied, risk modellers conduct studies on the wealth of the economy and then compute the aggregate expected catastrophe losses, $\mu_{\lambda_{R}W}$. Each insurer $j$ keeps a reserve amount per policy supplied as follows:
\begin{equation}\label{eq:reserve}
\text{reserve per policy}=F^{-1}(\rho_j)
\end{equation}
where $F$ is the cdf of $N(\mu_{\lambda_{R}W}, \sigma_{\lambda_{R}W})$ and $\rho_j$ is the solvency percentile of insurer $j$ (see Section \ref{inssetup} for a recap on insurer setup). 

Only at time $0$, each insurer $j$ pays a fixed cost to the risk-modeller for their services in providing a loss model described above. At subsequent time-steps, insurers update their loss model based on the state of the environment. While no catastrophes occur, each insurer $j$  updates the reserve requirement per policy supplied to reflect changes in societal expected wealth loss using Equation (\ref{eq:reserve}).

\subsubsection*{Response to Catastrophes}
If a catastrophe occurs, however, each insurer $j$ responds by raising their catastrophe loss probability, $p_j$, by their catastrophe bias $\beta'_j$, Equation (\ref{eq:p}). Additionally, each insurer $j$ raises their solvency percentile $\rho_j$, Equation (\ref{eq:rho}), as a reflex response to catastrophes. This will serve to restrict the supply of insurance policies and raise their premium rates as noted in \cite{kunreuther2013insurance} and \cite{smetters2008financing}. Another attribute that responds to catastrophe experience is the insurer's proportion of assets available for insurance, $\gamma_j$, which decreases after a catastrophe experience, Equation (\ref{eq:gamma}). This means that insurers are willing to dedicate less of the total assets toward insurance after a catastrophe experience.

\begin{equation}\label{eq:p}
    p_{j}=p_{j}(1+\beta'_j)
\end{equation}
\begin{equation}\label{eq:rho}
    \rho_{j}=\rho_{j}(1+\beta'_j)
\end{equation}
\begin{equation}\label{eq:gamma}
    \gamma_{j}=\gamma_{j}(1-\beta'_j)
\end{equation}

\subsubsection{Deriving Catastrophe Insurance Supply}
\label{supply}
Given a chosen loss model and the initial level of capital $\kappa_0$, each insurer $j$'s total assets at any time-step $t$ is summation of all its accounts:
\begin{equation}\label{eq:assets}
    \text{assets}=\kappa_{tj}+ \text{profits}+ \text{reserves}+\text{reinsurance}
\end{equation}
In determining how much to supply in a given year, insurers use their disposable assets proportion $\gamma_j$, introduced in Section \ref{inssetup}, to determine their insurance supply. Insurance supply is hence calculated as:
\begin{equation}\label{eq:supply}
     \text{supply}=\frac{\gamma_j\text{(assets)}}{\text{reserve per policy}}
\end{equation}

Upon selling a catastrophe insurance policy, the insurer collects a premium amount, in each time-step in which the individual is insured, as follows:
\begin{equation}\label{eq:premium}
    \text{premium paid}=p_j(1+l_j)(\lambda_{Ri}W_{ti})
\end{equation}
 where $p_j$ is the loss probability or the risk premium rate and $l_j$ is insurer $j$'s premium loading, and $\lambda_{Ri}W_{ti}$ is claim amount paid if a catastrophe occured (see Section \ref{popsetup} for a recap on how individuals lost wealth due to catastrophes).
 
The expected profits from the sale of a policy is computed as:
\begin{equation}\label{eq:profit}
    E(\pi)=p_j(1+l_j)X-p_jX-c_j'p_jX 
\end{equation}
this is where $c'$ represents the administrative costs paid as a rate of the claim $X=\lambda_{Ri}W_{ti}$. Given Equation (\ref{eq:profit}), insurers maximize profits by setting the premium loading ($l_j$) as high as possible. However, loaded premium rates cannot persistently increase due to demand-side limitations and competitive pressures.

\subsubsection{Market Entry and Exit}
\label{entry}
In our environment, insurers can enter the market and start supplying catastrophe insurance policies as long as existing insurers make profits i.e. there are opportunities in the market. Entry will cease when all the existing insurers make zero-profits, i.e. equilibrium is achieved. On the other hand, if existing insurers make losses, they start exiting the market until incumbent firms make zero profits. 

\textbf{The exit parameter:}

The insurer exits the market once the exit parameter, $\epsilon_j$, reaches a value of 1 (see Section \ref{inssetup} for a recap on insurer setup). The exit parameter is incremented if any of the following events occur: (1) the insurer experiences losses in a given year, (2) the insurer experiences no sales in a given year, and (3) insurers constantly need to draw from their capital to fulfill claim liabilities and the reserves are not enough. 

Nonetheless, an insurer can immediately exit the market, regardless of the value $\epsilon_j$, if at any point in an episode the insurer becomes insolvent and is unable to pay for claim liabilities.

  \section{The Government Agent}
  \label{gov}

  In this section, we outline how we modeled our government agent. As a reminder to the reader, the government agent is the only RL agent in the environment. This means that it will be modeled to ``learn'' its optimal behavior i.e. intervention by iterative ``trial and error'' based on ``feedback'' it receives from the environment. This approach is contrasted with how we modeled the individual and the insurer agents in Sections \ref{ind} and \ref{ins} respectively. In the latter cases, we assumed underlying microeconomic models dictating agents' behavior. The government agent can be compared to a \textit{multiarmed bandit} agent where each of its arms correspond to a possible intervention policy in the environment.

  In the environment, the government agent can take three actions. First, it can \textit{intervene} in the catastrophe insurance market. Second, it can \textit{observe the welfare impact} of its intervention policy. Third, it can \textit{collect taxes} to finance its intervention policies. In this section, we will discuss how we modelled the government agent to take its actions. We assumed that the government perceived the true catastrophe risk in the environment $\theta$ and each individual's true wealth loss rate due to catastrophes $\lambda_R$.
  
  \subsection{Intervening}
  \label{intervene}
  At any time-step $t$ of the episode length, the government intervened from the following set of allowable interventions. These relate to a bandit's \textit{arms} in a multi-armed bandit context.

    \noindent An intervention at time $t$ took the form of a policy change as described below. These interventions are based on the literature we described in Section \ref{intro} and summarized in Table \ref{tab:pol}.
  \begin{itemize}
      \item Take no action: If the government believes that the environment is well-functioning and additional intervention would not improve welfare impact, it may choose to take no action. We added this intervention to ensure that it is an option for the government not to intervene at all if it wants to.
      \item Offer a government-provided insurance policy: 
      \begin{itemize}
          \item Intervention Description: This intervention is based on the discussion in \cite{Jaffee_Russell_2013}, \cite{kunreuther2013insurance},and \cite{smetters2008financing} on the effectiveness of state-provided insurance in the management of catastrophe risk. If the government chose to pursue this intervention policy, it would extend its provision of state-provided insurance to \textit{one} other individual. Government-provided insurance would be offered to the neediest individual, i.e. individual who would be  willing to pay the maximum premium for an insurance contract. The neediest individual can occasionally be policyholder under a private insurer. If the latter is the case, the individual becomes the neediest if the difference between the individual's $P^{max}_{ti}$ and the amount they currently pay their private insurer exceeds the $P^{max}_{ti}$ of any other individual. 
          \item Premium Rates: Once identified, the neediest individual would become a government policyholder paying a premium rate that is the minimum between the fair premium rate, $\theta\lambda_{Ri}W_{ti}$, and the maximum premium the individual is willing to pay $P^{max}_{ti}$. If policyholder pays under the latter option, the government finances the deficit from taxpayers' funds (see Section \ref{taxsystem} for details on government's tax system). As long as an individual is a government-provided policyholder, premiums payable are updated at each time period $t$ for the changes in individual's wealth, catastrophe wealth loss rates and other parameters.
          \item Insurance Period: An individual would remain government-insured at any time-period $t$ as long as there are no cheaper offers from private insurers. If there are cheaper private insurers, the individual cancels the government insurance contract and is no longer eligible for government insurance even if they become uninsured at a later time-period.
      \end{itemize}
      \item Ease insurer solvency requirements: This intervention is based on discussion in \cite{Jaffee_Russell_2013} and \cite{cummins2002can} who questioned insurers' ability to handle the huge losses that come with insuring catastrophes. Through this intervention, the government reduces each insurer's $j$ required reserve per policy, $\rho_j$. However, the $\rho_j$ would not fall below 70\%. This means, with the application of this intervention, no insurer would be required to keep in reserves less than the \nth{70} percentile of the expected losses of his clients (see Section \ref{lossmodel} for details on insurers' solvency requirements).
      \item Provide early warnings/awareness campaigns: This intervention is based on a possible remedy described in \cite{meyer2017ostrich}, \cite{kunreuther2013insurance}, and \cite{smetters2008financing}. If the government chose to pursue this policy intervention, it would educate each individual $i$ on the true catastrophe risk, $\theta$, and the true catastrophe wealth loss rate, $\lambda_{Ri}$. The effectiveness of this intervention varied depending on when it was applied. When applied for the first time, the intervention has maximum effectiveness in adjusting individuals' perception of catastrophe risk and expected wealth loss due to catastrophes. However, a \textit{cry-wolf} effect happens when the government kept pursuing this policy without observing a real catastrophe experience that supports the government's warnings (see Section \ref{cost} for details on the cry-wolf effect).
      \item Increase subsidies on insurance policy premiums: This intervention is based on the recommendations described in \cite{kunreuther2013insurance} and \cite{smetters2008financing}. Under this policy intervention, the government would pay for a certain share, $s$, of a poliycholder's payable premium. When pursued, this intervention applied to all actual and prospective policyholders in the environment.
      \item Increase premium regulations: This intervention is based on the recommendations described in \cite{kunreuther2013insurance} on ensuring that insurance premium rates reflect the true likelihood of catastrophe in society. Under this intervention, the government would impose a premium rate ceil, $p^{reg}$, on private insurers. It also ensures that all private insurers are charged a fair loss premium rate and imposed ceils on premium loadings. 
      \item Offer a disaster prevention method: 
      This intervention is inspired by the recommendation presented in \cite{meyer2017ostrich} and \cite{smetters2008financing}. If the government chose to pursue this intervention, it would offer \textit{one} free disaster prevention method to the individual who would need it the most. The disaster prevention method reduces the treated individual's true and perceived catastrophe wealth loss. The neediest individual would be identified as the one who has the highest true catastrophe wealth loss rate, $\lambda_{Ri}$. 
         
      \item Offer government reinsurance: This recommendation is based on the policy prescriptions provided in \cite{cummins2002can}, \cite{bruggeman2010government}, \cite{kunreuther2013insurance} where the government acts as a reinsurer of last resort. This intervention policy, when pursued, enables private insurers to get reinsurance contracts from the government at a government-set reinsurance rate. These contracts serve to cover the losses of a private insurer at the risk of defaulting. The intervention policy also allows the government to act as a re-insurer of last resort when a private insurer, be a reinsurance contract holder or not, fails to meet its claim liabilities after a catastrophe. 
  \end{itemize}
  \subsection{Observing Welfare Impact of Policy Intervention}
  \label{observe}
  After choosing a policy intervention, the government observed its welfare impact for future reference. There are many ways a government could compute the welfare impact of a policy intervention. For our purposes, we adopted the Marginal Value of Public Funds ($MVPF$) approach described in \cite{finkelstein2020welfare} and \cite{hendren2020unified}. Similar to social cost-benefit approaches, the $MVPF$ is computed as:
\begin{equation}\label{eq:45}
    MVPF=\frac{\text{Willingness to pay for policy by beneficiaries ($WTP$)}}{\text{Net cost of Government Spending ($G$)}}
\end{equation}
Relying on the $MVPF$ was particularly relevant for our purposes as it allowed us to measure the long-term impact of an intervention policy. This feature made it a strong fit to act as a reward function in the Q-learning algorithm we would be using to allow the government to learn its optimal interventions under various states or 'contexts' of our environment. By design, the Q-learning algorithm maximized the long-run expected reward (see Section \ref{qlearning} for details on the Q-learning algorithm).

We set up the $MVPF$ framework by identifying the WTP and the G of each policy intervention outlined in Section \ref{intervene} above.
\subsubsection{Willingness to Pay (WTP)}
\label{wtp}
We compute the willingness to pay of the intervention policy changes, described above, as follows. Our equations take into consideration the notations we introduced in Sections \ref{setup}, \ref{popsetup}, \ref{inssetup}.
 
 \begin{itemize}
    \item Take no action: This policy change has no effect on individuals. Therefore, the WTP of this policy change is always 0.
     \item Offering government-provided insurance: The marginal individual, who is selected by the government to receive government insurance, is the one who would be willing to pay for this policy change. The WTP of the marginal individual depends on whether they were already a policyholder or not at time $t$. If the marginal individual was a policyholder, they would not need to switch to government insurance unless it is cheaper. Therefore, their WTP would be the difference between the premium they are paying for their existing policy, $p_j\lambda_pW_{ti}$, and the premium offered by government insurance policies, $g\lambda_pW_{ti}$. If the marginal individual is not a policyholder, their WTP would be the maximum amount they would be willing to pay for catastrophe insurance i.e. $P^{max}_{ti}$. Therefore, the WTP for this policy change can be expressed as:
     \begin{equation}\label{eq:WTPpol1}
         WTP_1= H_0P^{max}_{ti}+(1-H_0)(p-g)(\lambda_pW_{ti})
     \end{equation}
    where $H_0$ refers to the indicator variable that indicates whether the marginal individual was a policyholder $(H_0=0)$ or not $(H_0=1)$.
     \item Easing Insurer Solvency Requirements: Easing solvency requirements would reduce the amount private insurers need to keep in reserves per policy supplied. This will enable private insurers to supply more catastrophe insurance policies. Non-policyholders at time $t$ who purchase private insurance due to this policy change would be willing to pay, at most, $P^{max}_{ti}$.  Therefore, we express the WTP for this policy change as sum of the $P^{max}_{ti}$s of the marginal individuals who would purchase from the marginal catastrophe insurance policies supplied by insurers in response to easing solvency requirements:
     \begin{equation}\label{eq:wtppol2}
         WTP_2=\sum H_2P^{max}_{ti}
     \end{equation}
     where $H_2$ refers to the indicator variable that indicates whether the individual $i$ was a marginal individual.
     \item Provide early warnings/awareness campaigns: We assumed that this policy change affected all individuals in the environment. The WTP for any individual $i$  is the difference between the maximum  premium they would pay under the effects of underestimated risk perceptions, $P^{max}_{ti}$, and the true payable premium under the risk perceptions corrected by the policy, $P^{T'}_{ti}$. Therefore, the $WTP$ can be expressed as:
     \begin{equation}\label{eq:wtppol3}
         WTP_3=\sum (P^{T'}_{ti} -P^{max}_{ti})
     \end{equation}
     \item Increase subsidies on insurance policy premiums: This policy change affects two groups of individuals: (1) existing policyholders and (2) marginal non-policyholders who would become policyholders after applying subsidies to premium rates. For the first group, the WTP would be the premium amount they no longer need to pay due to subsidies, i.e. $(p-s)\lambda_PW_{ti}$. For the second group, the WTP would be the difference between the maximum amount they would be willing to pay for insurance and the subsidized premium they end up paying, i.e. $P^{max}_{ti}-(1-s)p\lambda_PW_{ti}$ This rationale can be summarized in the expression below:
     \begin{equation}
         WTP_5=\sum(H_0(P^{max}_{ti}-(1-s)p\lambda_PW_{ti})+(1-H_0)(p-s)\lambda_PW_{ti})
     \end{equation}
     where $H_0$ refers to the same indicator variable used in Equation (\ref{eq:WTPpol1}).
     
      \item Increase premium regulations: In a similar construction to $WTP_5$, this policy change affects current policyholders at time $t$ and marginal non-policyholders who become policyholders at time-step $t+1$. Individuals who were policyholders at both time $t$ and $t+1$ would have a WTP that is equal to the difference between the premium paid pre and post regulation. Marginal individuals who would purchase an insurance policy post premium regulation would have a WTP that is equal to the difference between the maximum premium they would be willing to pay and would they pay in their insurance contract. Formally, the $WTP$ can be expressed as:
     \begin{equation}\label{eq:wtppol6}
         WTP_6=\sum(H_0(P^{max}_{ti}-p^{reg}\lambda_PW_{ti})+(1-H_0)(p-p^{reg})\lambda_PW_{ti})
     \end{equation}
    where $H_0$ refers to the same indicator variables used in Equation (\ref{eq:WTPpol1}).
     \item Offering Disaster Prevention Methods: This policy change only affects the individual who would receive the disaster prevention method. The WTP of this marginal individual would be the amount saved, if a catastrophe happened, due to reducing the ex ante catastrophe wealth loss rate, $\lambda_R$, to the ex post catastrophe wealth loss rate, $\lambda^D_R$. Therefore, the WTP can be expressed as:
     \begin{equation}
         WTP_7= (\lambda_{Rti}-\lambda^{D}_{Rti})\alpha_{ti}W_{ti}
     \end{equation}
     This is where $\lambda^{D}_{Rti}$ is the real loss incurred by an individual $i$ due to catastrophes after investing in a disaster prevention method.
     
     \item Increase Reinsurance Funds: Only individuals who are policyholders both at times $t$ and $t+1$ would be willing to pay for this policy change. If the insurer's exit parameter, $\epsilon$, denotes their insolvency risk, policyholders would be willing to insure against their claim default risk \textit{if} a catastrophe happened. Formally, $WTP$ can be expressed as:
     \begin{equation}\label{eq:WTPpol8}
         WTP_8=\sum ((1-H_0)*(\epsilon_j\alpha_{ti}(\lambda_pW_{ti})))
     \end{equation}
     where $H_0$ refers to the same indicator variable used in Equation (\ref{eq:WTPpol1}).
 \end{itemize}
  \subsubsection{Net cost of Government Spending (G)}
  \label{cost}
  As described by \cite{finkelstein2020welfare}, the Net Cost to Government component of the $MVPF$ is composed of: (1) mechanical cost of policy change and (2) fiscal externalities. In our model, the mechanical cost is financed through taxation (see Section \ref{taxsystem} for the government's tax system). The fiscal externalities are unintended effects that occur due to or in spite of policy changes. These are welfare costs/benefits incurred due to a policy change and for which no one pays for.
  
  \noindent The estimated mechanical cost of each policy change is described in the list below. 
\begin{itemize}
\item No Action: This policy change is costless to the government.
    \item Offering government-provided insurance: We assumed that the mechanical cost of offering government-provided insurance policies is a fixed value $x$ per policy. This corresponds to the potential administrative and marketing costs that came with the insurance. 
    \item Easing Insurer Solvency Requirements: We assume that this policy change costs a fixed amount $x$ of administrative costs needed impose this change on private insurers. 
    \item Provide early warnings/awareness campaigns: The mechanical cost of educating an individual $i$ about their risk exposure depends on their social class, $c$, (see Section \ref{ind} for a recap on individuals' attributes). It costs a fixed amount $x$ to correct the beliefs of a middle class individual. It costs a fixed amount $1.5x$ ($0.5x$) to correct the beliefs of a low (upper) class individual.
    \item Increase Government Subsidies on Premium Rates: If the government intends to reduce the premium rates of private insurers by a share $s\in [0,1]$ through subsidies, the cost of the subsidy per individual $i$ of getting insurance from insurer $j$ is $s*p_j\lambda_{Pti}W_{ti}$.
    
    \item Insurance premium regulation: The mechanical cost of imposing premium regulations is a value $x$ representing the required administrative costs to impose this change on existing private insurers.
    
     \item Offering Disaster Prevention Methods: The mechanical cost of this policy change is an amount $x$ for every unit of wealth at risk reduced by disaster prevention methods, $(\lambda_{Rti}-\lambda^D_{Rti})W_{ti}$.
    
    \item Increase Reinsurance Funds: The mechanical cost of this policy change is the amount each insurer expects it would need over and above its current total assets.
   
\end{itemize}

\noindent In addition to the mechanical cost, the allowable policy interventions above can be associated with the fiscal externalities below:
\begin{enumerate}
    \item Crowding Out Private Insurers: If the government charges insurance at lower premium rates than private insurers, individuals are likely to purchase insurance policies from the government. This can crowd out private insurers. We measure this fiscal externality as the sum of the total assets available with a crowded-out insurer. This represents the amount an insurer could have used to supply catastrophe insurance policies but was lost instead.
    \item Debt-related Tax Raises: If at any time $t$ the government is required to pay more claim liabilities due to a catastrophe than it has in its funds, it is obliged to borrow money. To pay its interest-bearing debts, the government would be obliged to raise taxes on individuals in the future. We measure costs related to this externality as the unexpected tax raises intended to serve public debt.
    \item Private Insurer Insolvency: If the sequence of government interventions until time-step $t$ failed to prevent insurers from becoming insolvent, individuals can be left vulnerable against catastrophe losses. We measure this fiscal externality as the sum of the unmet liabilities insurers were not able to pay due to their insolvency. 
    \item Cry-Wolf Effect: This externality represents individuals' reduced receptiveness to educational campaigns and early warnings when they fail to see real-evidence of catastrophes. The cry-wolf effect was described a potential caveat in the effectiveness of educational campaigns in \cite{smetters2008financing}. We measure this fiscal externality by measuring an individual's expected risk exposure that went unperceived due to the cry-wolf effect of educational campaigns.
    \item Moral Hazard: An undesirable consequence of government intervention happens when it changes individuals' behavior making them more exposed to catastrophe risk, for example by building property in high-risk areas. This change in intentions due to government policy change was described in \cite{smetters2008financing} and \cite{kunreuther2013insurance}. We measure this fiscal externality by measuring the deviation between pre-intervention risk exposure and post-intervention risk exposure at a constant level of wealth.
    \item Catastrophe-related losses: If a catastrophe hits the environment, the amount of wealth lost is a loss incurred by society that is captured as a fiscal externality. It indicates government policy failure in preventing wealth losses due to catastrophes on society.
    \end{enumerate}

A single policy change may trigger one or more of the above fiscal externalities. Through repeated training  over a large number of unique episodes, the government agent would estimate the fiscal externality related to a policy change (see Section \ref{qlearning} for details on the Q-learning algorithm and Section \ref{results} for the $MVPF$ estimates of policy changes). 
  \subsection{Collecting Taxes}
  \label{taxsystem}
 At any time-step $t$, the government has to make certain payments including premium subsidies, catastrophe loss claim payments, and surprise reinsurance. To finance these payments, we assumed that the government adopts a progressive income tax system which raised funds from the individuals in the society.
 
 In its tax system, the government's objective is to raise funds that are enough to meet its liabilities yet are socially fair. The progressive tax system divided the individuals in the society into three groups: low, middle, and high income groups. Each individual $i$ in the society would be assigned to a tax rate based on the income group it belonged to and would pay its tax dues out of its constant income, $Y_i$, at any time-step $t$. (See Section \ref{popsetup} for a recap on individual attributes and assumptions related to them.)
 
 \section{Testing the Environment: Allowing No Government Intervention}
 \label{nogov}
 In this section, we evaluate the extent toward which the environment we developed in Sections \ref{envrionment}, \ref{ind}, \ref{ins} is reflective of stylized facts of real free catastrophe insurance markets i.e. those operating \textit{without government intervention}. We parameterized the attributes defined in Sections \ref{popsetup} and \ref{inssetup} by assuming distributions for them in our environment (see Appendix \ref{parameters} for details on the parameterization of the environment's attributes). We, then, run several episodes of our environment (see Section \ref{timeline} for a recap on episode timeline). We implemented the environment on Python and the assignment of the parameters is discussed in Appendix \ref{parameters}.\footnotemark \footnotetext{A link with the full running code of our model will be provided in the final submission of this paper.}

\subsection{Stylized facts of free catastrophe insurance markets}
\label{numtrend}
According to literature, catastrophe insurance markets exhibit certain characteristics found across a large sample of real catastrophe insurance markets (see Section \ref{intro} for a recap on real examples of failures of free catastrophe insurance markets). We will evaluate the extent towards which our environment is reflective of the following stylized facts:
\begin{enumerate}
    \item Free catastrophe insurance markets generate inadequate coverages \citep{charpentier2014natural}.
    \item Purchases of catastrophe insurance tend to increase right after the society witnesses a catastrophe experience and while the memory is still there \citep{kunreuther2013insurance}, \citep{gallagher2014learning}, \citep{atreya2015drives}, and \citep{dumm2020representative}.
    \item Policyholders tend to cancel their insurance coverages as the memory of past catastrophe fades and the time since the last catastrophe exceeds 5 years \citep{kunreuther2013insurance} and \citep{atreya2015drives}.
    \item After a catastrophe experience, many insurers respond by restricting supply and raising premium rates \citep{kunreuther2013insurance} and \citep{smetters2008financing}.
    \item Many insurers exit the market right after a catastrophe experience. For example, many private insurers cut their terrorism insurance supply right after 9/11 attacks \citep{cummins2002can} and \citep{kunreuther2013insurance}.
    \item Many individuals are willing to purchase catastrophe insurance but they are unable to find insurance at an affordable price \citep{kousky2012explaining}.
\end{enumerate}

\subsection{The Environment in Graphs{\footnotemark}}\footnotetext{All figures in this paper are best analyzed in a colored version of this submission. We strongly recommend the reader to read a colored version of the paper.}
In the subsequent tests, we assumed that our environment was composed of a population of $n=100$ individuals and $m=5$ initial insurers (see Sections \ref{ind} and \ref{ins} for a recap on the individual and insurer agents). An important assumption we also made was that poorer individuals of the environment were more likely to lose more as a proportion of their wealth due to catastrophes, i.e. poorer individuals have a higher $\lambda_{R}$, than richer individuals. We base this assumption on multiple studies which suggested that poorer individuals are likely to be exposed to greater catastrophe risk due to locating in high risk prone areas, and lower re-location abilities \citep{perlin2001residential}, \citep{boustan2012moving}, \citep{boustan2020effect}, and \citep{SMAHSA2020}.

\subsubsection{Proportion of covered losses}
\label{coveredloss}
Without government intervention, we are interested in observing the coverage rate in our environment. We defined the coverage rate as the sum of the expected loss due to catastrophes as a proportion of total expected losses in the environment. Below we document two types of behavior exhibited by selected individuals in our environment. We observed the coverage rate under two different scenarios: (1) no catastrophe experience at all, and (2) more than one catastrophe experience. 
\begin{figure}[!ht]
  \centering
  \subfloat[No Catastrophe Experience]{\includegraphics[width=0.4\textwidth]{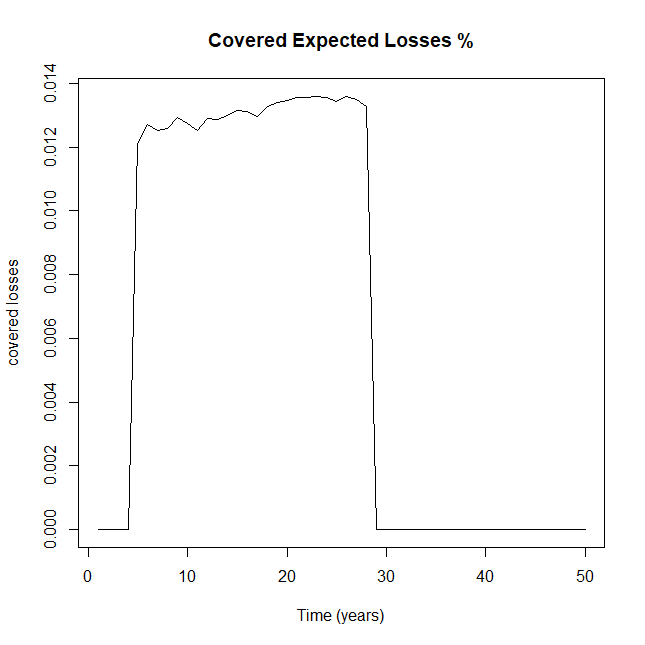}}
  \hfill
  \subfloat[2 Catastrophe Experiences (Red Lines)]{\includegraphics[width=0.4\textwidth]{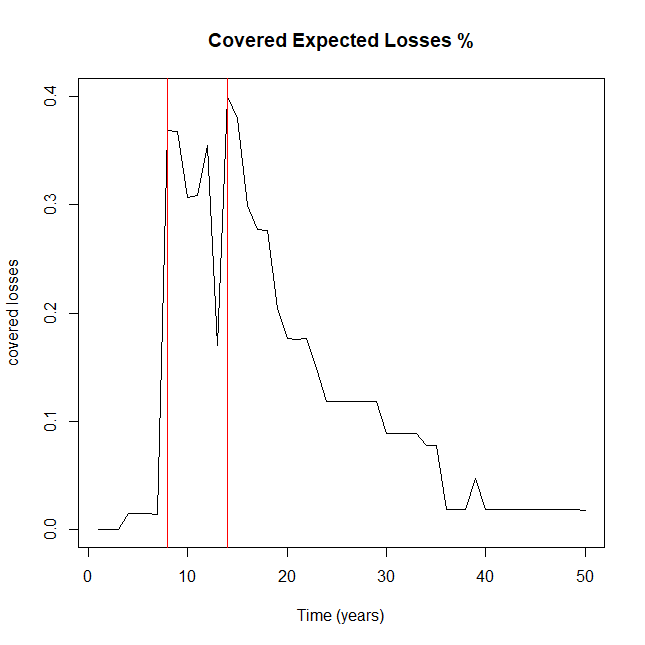}}
  \caption{(a) Covered expected losses as a proportion of total losses show a short-term rise in coverage but falls again to 0\% by the end of the episode. Moreover, the maximum coverage percentage in the episode is little over 1\%, which is inadequate and consistent with stylised fact 1 discussed in Section \ref{numtrend}. (b) Covered expected losses as a proportion of total losses increases with catastrophe experience (red line) but declines as the memory of catastrophe fades with time, consistent with trend 2 in Section \ref{numtrend}. However,  the maximum covered percentage is less than 40\% and the episode ends with less than 1\% coverage which is inadequate and consistent with trend 1 in Section \ref{numtrend}.}
\end{figure}
\FloatBarrier
\subsubsection{Evolution of Gini Index}
We explored what happens to the Gini Index in the our environment with and without catastrophe experience. In those experiments, we temporarily assumed that no individual purchased catastrophe insurance at any time-point in the episode. We made this assumption to observe the full effect of catastrophes on wealth equality in the society, without being distorted by those individuals who protect their wealth. We found that the Gini index rises with each catastrophe experience, consistent with studies on the effects of catastrophes on income and wealth distribution \citep{boustan2020effect}. 
\begin{figure}[!ht]
  \centering
  \subfloat[No Catastrophe Experience]{\includegraphics[width=0.4\textwidth]{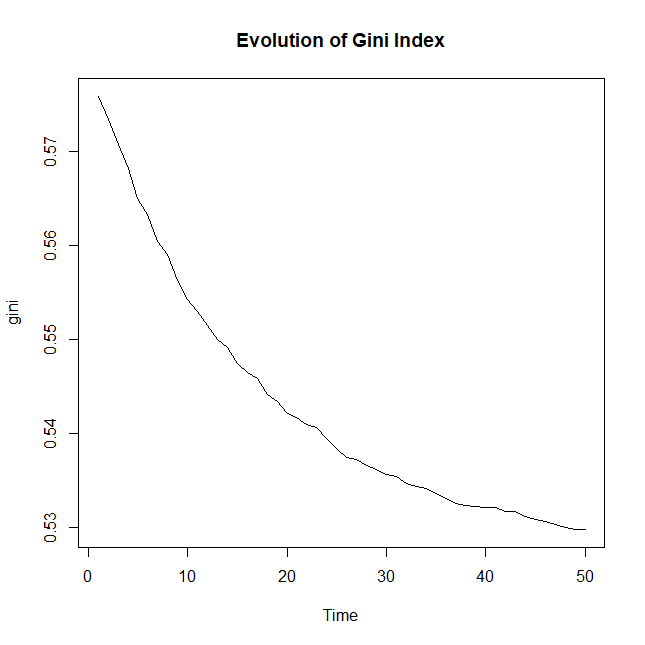}}
  \hfill
  \subfloat[5 Catastrophe Experiences]{\includegraphics[width=0.4\textwidth]{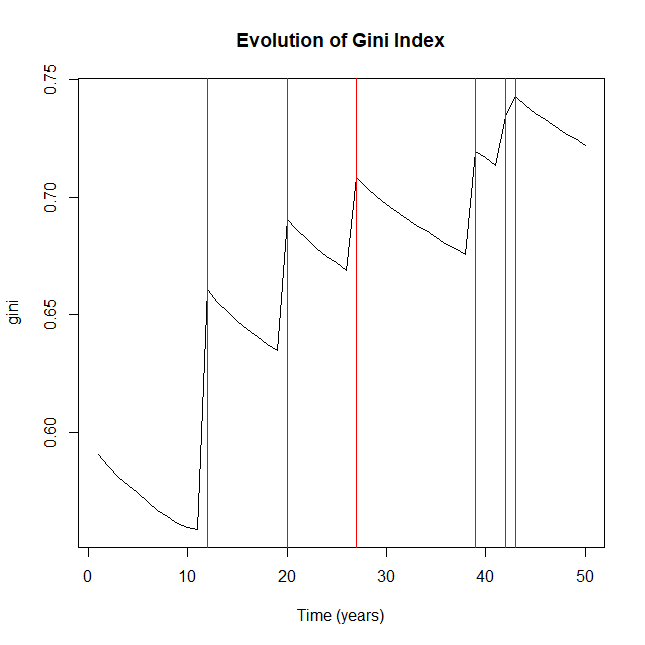}}
  \caption{(a) Gini index declines as all individuals uninterruptedly accumulate savings. (b) Gini index hikes after catastrophe experience (red lines) because catastrophes cause greater losses on the poorer segments of society than the richer segments.}
\end{figure}
\FloatBarrier

\subsubsection{Individuals' Insurance Purchase Behavior}
 From our experiments, we present three dominant types of individual agent behavior. The heterogeneity in the behavior of the individual agent was achieved by heterogeneous initialization of their attributes of  (see Appendix \ref{parameters} for agent attribute assignment). We present behavior from Episode A in which individuals experienced no catastrophes and Episode B in which individuals experienced two catastrophe experiences.
\begin{figure}[!ht]
  \centering
  \subfloat[Episode A: Risk Perception ]{\includegraphics[width=0.5\textwidth]{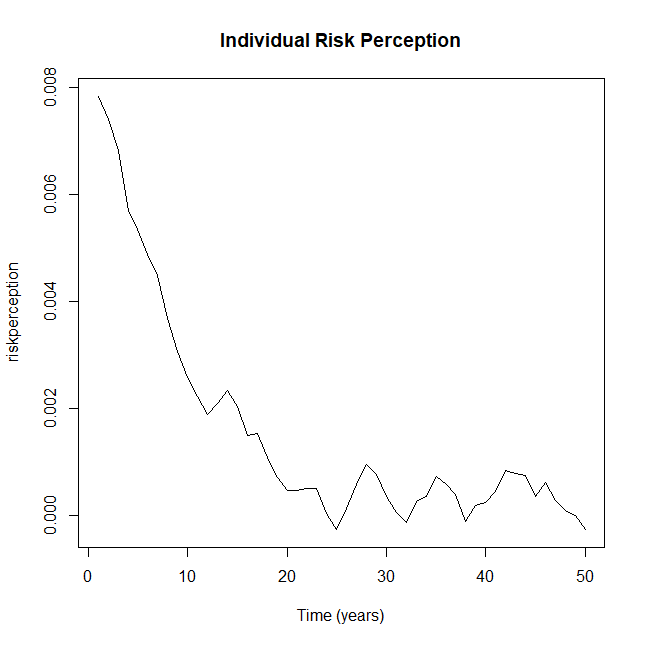}}
  \hfill
  \subfloat[Episode A: Wealth ]{\includegraphics[width=0.5\textwidth]{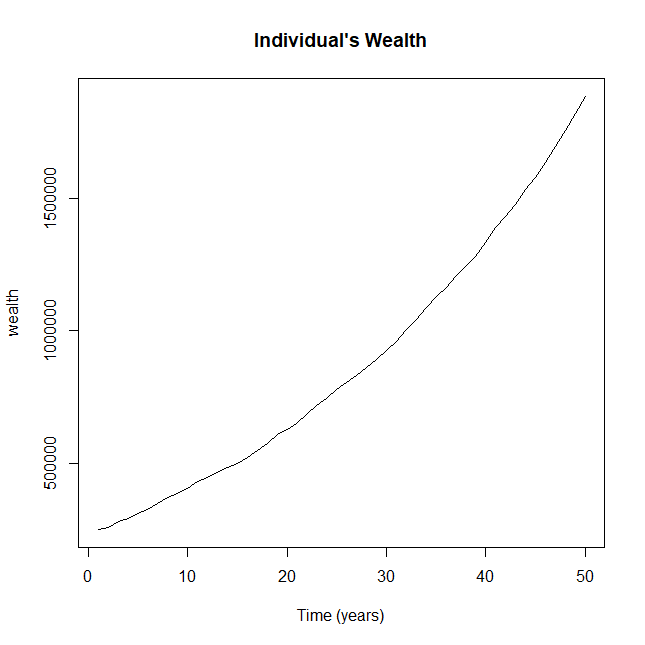}}
  \hfill
  \subfloat[Episode A: Premium Demanded VS Premium Offered ]{\includegraphics[width=0.5\textwidth]{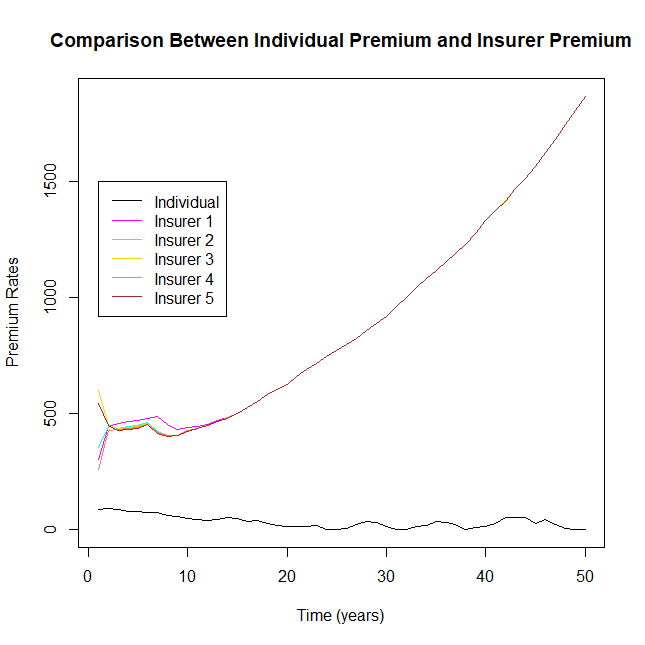}}
  \caption{\small (a) Individual's risk perception declines with no evidence of catastrophes occurring, this decline is sustained by individuals' optimism and amnesia biases, (b) Individual's wealth constantly increases as they accumulate savings without catastrophe disruptions, (c) Individual's $P^{\text{max}}$ is far below any premium rate offered by the available insurers in the market. This makes insurance purchase unfeasible and is consistent with stylised fact 6 in Section \ref{numtrend}. Overall, when no catastrophes happen, very few individuals purchase insurance at any point in Episode A.}
\end{figure}
\FloatBarrier
\begin{figure}[!ht]
  \centering
  \subfloat[Episode B: Risk Perception ]{\includegraphics[width=0.5\textwidth]{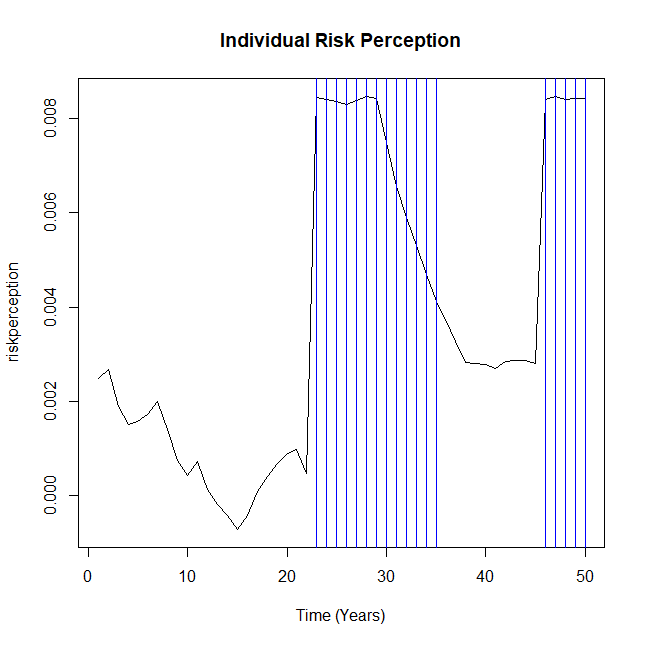}}
  \hfill
  \subfloat[Episode B: Wealth]{\includegraphics[width=0.5\textwidth]{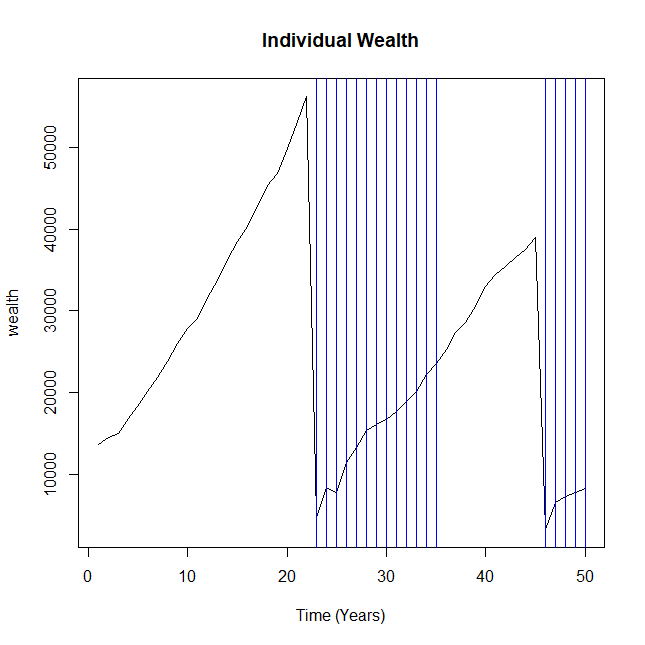}}
  \hfill
  \subfloat[Episode B: Premium Demanded VS Premium Offered ]{\includegraphics[width=0.5\textwidth]{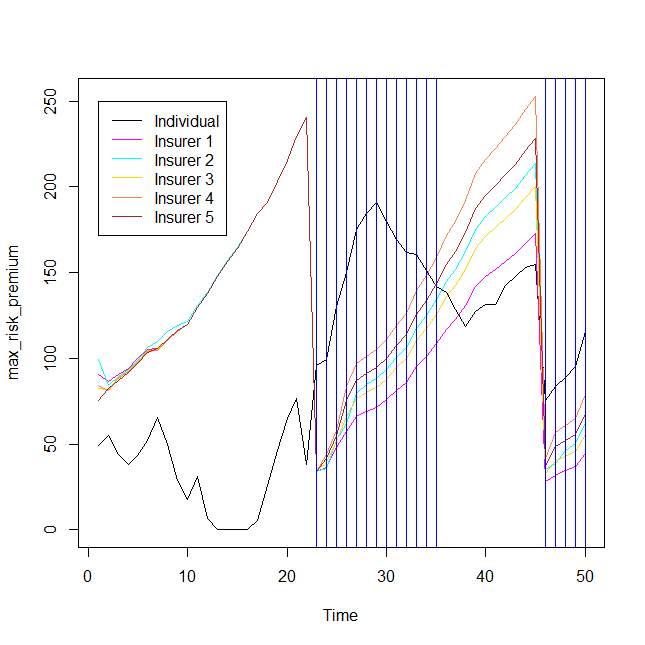}}
  \caption{(a) Individual risk perception spikes after catastrophe experience, after which they immediately purchase insurance. They remain covered for the time period highlighted by the blue lines. As time since the last witnessed catastrophe increases, individual's risk perception falls until they cancel their insurance coverage. Another catastrophe occurs at time 45 and the individual's risk perception spikes again. The individual re-purchases insurance after the shock. (b) The individual was covered by catastrophe insurance (blue lines) after they witnessed a catastrophe experience, therefore, their wealth was always shocked by the catastrophes, (c) Individual's $P^{\text{max}}$ only exceeded insurers' premium rates after the catastrophe experience. However, it fell below the rates of all other insurers as risk perception fell. All these observations are consistent with stylised facts 2, 3, and 6 from Section \ref{numtrend}}
\end{figure}
\FloatBarrier
\begin{figure}[!ht]
  \centering
  \subfloat[Episode B: Risk Perception ]{\includegraphics[width=0.5\textwidth]{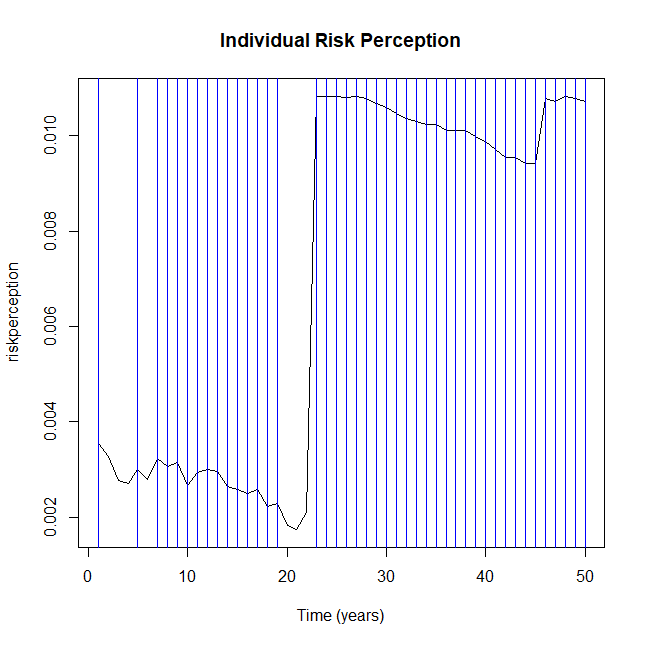}}
  \hfill
  \subfloat[Episode B: Wealth ]{\includegraphics[width=0.5\textwidth]{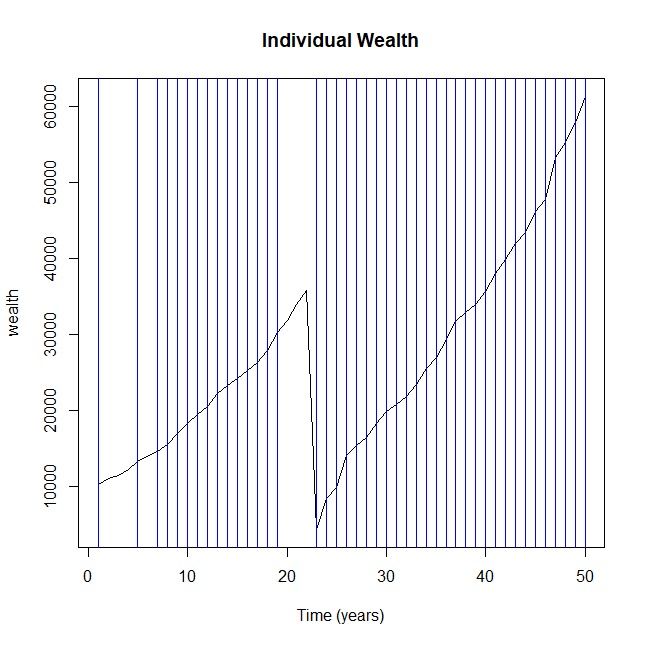}}
  \hfill
  \subfloat[Episode B: Premium Demanded VS Premium Offered ]{\includegraphics[width=0.5\textwidth]{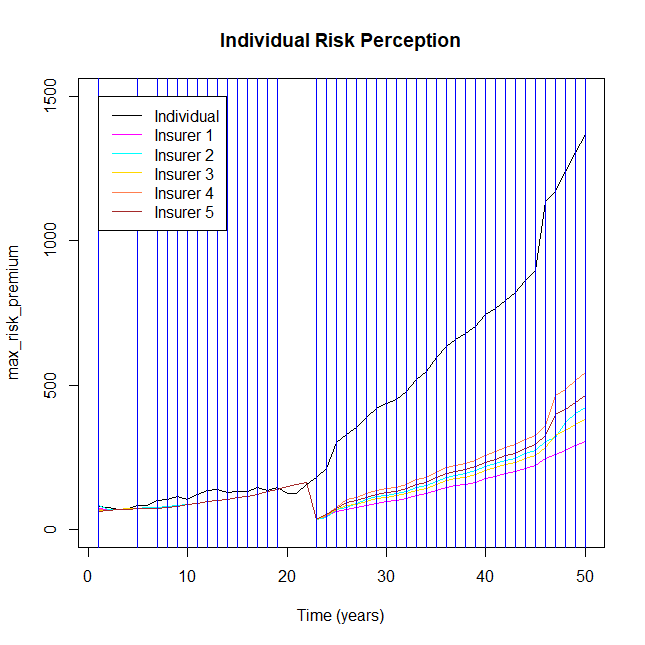}}
  \caption{(a) This individual's risk perception was relatively high enough, due to heterogeneity in the environment, to allow them to be covered (blue lines) before experiencing a catastrophe experience. However, they cancelled their catastrophe insurance coverage right before a catastrophe happened in society. After the catastrophe happened, their risk perception spiked and they repurchased insurance and remained covered until the end of the episode, (b) Individual's wealth is shocked by the first catastrophe as they were uninsured. After the first catastrophe, they remain insured throughout the episode and their wealth was unaffected by the second catastrophe as opposed to the individual in Figure 6, (c) The individual's pre-catastrophe $P^{\text{max}}$ was higher than some of the active insurers. Post-catastrophe, it remained higher than the rates of all other insurers and this allows them to be insured throughout the entire episode. }
\end{figure}
\FloatBarrier
\subsubsection{Evolution of Insurer Premium Rates}
In  our experiments, we found that each catastrophe experience caused insurers to increase premium rates after the shock in an irrational behavior.
\begin{figure}[!ht]
  \centering
  \subfloat[No Catastrophe Experience]{\includegraphics[width=0.35\textwidth]{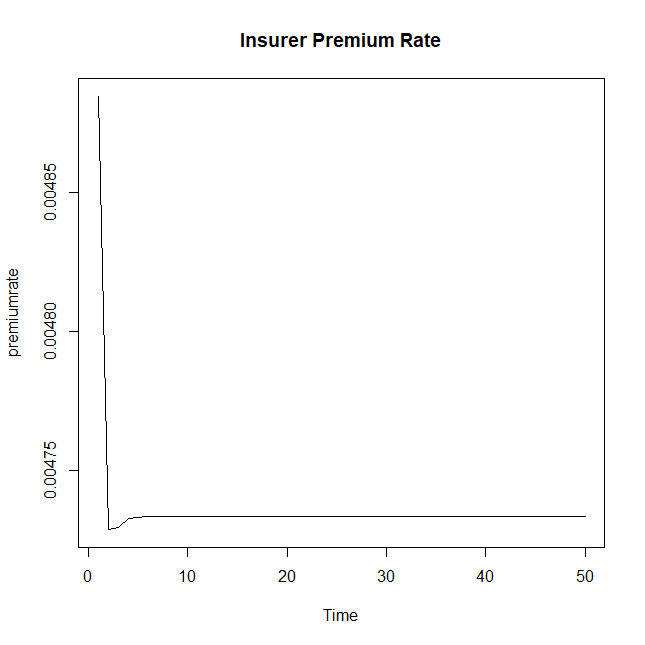}}
  \hfill
  \subfloat[5 Catastrophe Experiences]{\includegraphics[width=0.35\textwidth]{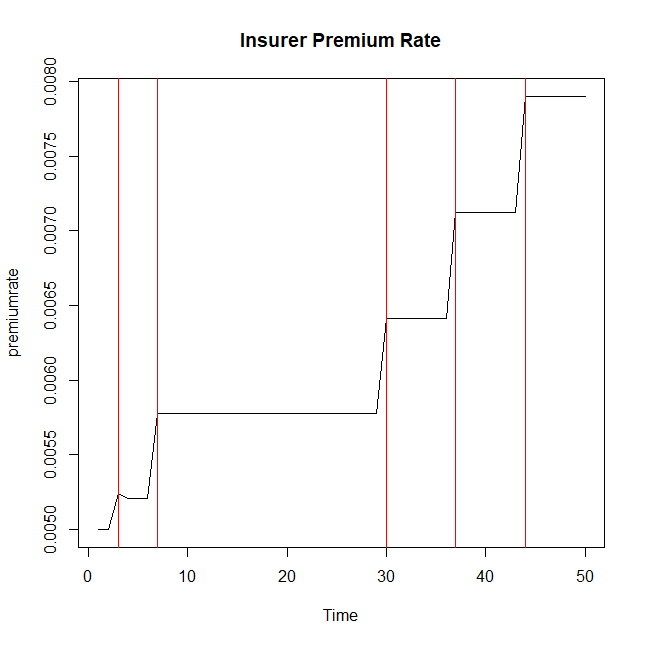}}
  \hfill
  \caption{ \small (a) The insurer premium rate declines as no sales happen and to coverage with the market's average premium rate by competitive pressures. The insurer eventually exits the market. (b) Insurers raise their premium rates, irrationally after catastrophe experiences, consistent with stylised facr 4 in Section \ref{numtrend}. }
\end{figure}
\FloatBarrier
  \section{Testing the Environment: Allowing Government Intervention}
  \label{qlearning}
  Up until this point in the paper, we designed an environment composed of three types of agents: individual, insurer, and government (see Section \ref{envrionment} for a recap on environment setup and Sections \ref{ind}, \ref{ins}, \ref{gov} for a recap on agent design). We also observed, in many  experiments, that the environment reflected many stylised facts about real catastrophe insurance markets without government intervention (see Section \ref{nogov} for a recap). In this section, we allow our government agent to intervene in the environment with the goal of maximizing long-term welfare of the individuals under catastrophe risk. We are particularly interested in observing which of the interventions have the highest $MVPF$, marginal value of public funds, under various situations (see Section \ref{intervene} for a recap on the government's allowable interventions and Section \ref{observe} for a discussion on policy $MVPF$). By reinforcement learning, the government agent will “learn”
its optimal behavior, i.e. intervention, by iterative “trial and error” based on “feedback” it receives from
the environment. The reinforcement learning algorithm we will use in this work is Q-learning.
  \subsection{Government Intervention: A Q-learning Problem}
  A Q-learning algorithm is a model-free\footnotemark \footnotetext{A model-free algorithm is one that does not rely on a transition probability matrix of the underlining environment to solve the problem of deriving the optimal policy intervention. This means model-free algorithms help the agent derive optimal policy with minimal knowledge of the environment.}  off-policy\footnotemark \footnotetext{An off-policy learner learns the value of optimal policy independently of the current action taken. This is contrasted with on-policy learning where the learner learns the value of the policy currently pursued.}  reinforcement learning algorithm that iteratively learns the quality of taking an action in a given state of the environment. A Q-learning problem is composed of a learning agent, an environment with which the learning agent interacts, an action space which the learning agent uses to intervene in the environment,  a reward function which provides the agent with feedback on the quality of the intervention, and a state space which is equivalent to the "context" in reference to contextualized bandits. 
  For our purposes:
  \begin{itemize}
      \item \textbf{Learning Agent:} The Government Agent designed in Section \ref{gov}
      \item \textbf{Envrionment:} The Environment in Section \ref{timeline}
      \item \textbf{Action Space:} The set of allowable policy interventions as defined in Section \ref{intervene}
      \item \textbf{Reward:} The Marginal Value of Public Funds (see Section \ref{observe} for recap)
      \item \textbf{State Space:} We look at the individuals' \textbf{awareness} of their catastrophe risk exposure, the \textbf{availability} and the \textbf{affordability} of catastrophe insurance policies in the market. We will precisely consider the following states:
      \begin{itemize}
      \item State 1: low awareness, low supply of catastrophe insurance, affordable premium rates.
      \item State 2: low awareness, high supply of catastrophe insurance, affordable premium rates.
      \item State 3: low awareness, high supply of catastrophe insurance, unaffordable premium rates.
      \item State 4: high awareness, low supply of catastrophe insurance, unaffordable premium rates.
      \item State 5: high awareness, high supply of catastrophe insurance, unaffordable premium rates.
  \end{itemize}
  \end{itemize}

\subsubsection*{Learning the Optimal Policy}
  Our government agent, described in Section \ref{gov}, does not have any prior knowledge about the environment it is intervening in. It intervenes in the environment by taking an action via \textbf{exploration} or \textbf{exploitation}. Under exploration, the agent randomly picks an action from the action space to learn more about the underlying environment. Under exploitation, the agent picks the action that has so-far yielded the highest reward in the training process. Exploration and exploitation are two conflicting objectives where the former ensures that the government agent knows sufficiently enough about the environment and the latter speeds up the learning process. In our model, we handle the exploration-exploitation trade-off through an $\epsilon$ greedy strategy i.e. we define the probability with which the agent is likely to explore or exploit.

  Once taken by either exploration or exploitation, the action causes the other agents, i.e. the individuals and insurers, to change behavior thus generating a feedback signal observed by the government. The feedback may be in the form of a reward (penalty) that positively (negatively) reinforces the action and makes it more (less) likely to be taken again under similar circumstances. Our aim is to learn the Q-values of all the government's interventions. We define the Q-values, after which Q-learning is named, as the expected reward of an intervention at a given state:
  \begin{equation}
    \label{eq:bellman}
    q(S_t,A_t)=E[R_{t+1}+\gamma q(S_{t+1},A_{t+1})\mid S_t, A_t]
\end{equation}
   where $S_t$, $A_t$, $R_t$ denote the environment's state, from the state space, and the government's action, from the action space, and the environment's reward, i.e. the MVPF, respectively.
   
   Since our reward function is the policy's marginal value of public funds, $MVPF$, Equation (\ref{eq:bellman}) can be rewritten as:
   \begin{equation}
    \label{eq:bellmanmv}
    q(S_t,A_t)=E[MVPF_{t+1}+\gamma q(S_{t+1},A_{t+1})\mid S_t, A_t]
\end{equation}
   Using \cite{sutton2018reinforcement}, Equation (\ref{eq:bellmanmv}) can be rewritten in an iterative fashion as follows:
   \begin{equation}
   \label{eq:bellmanit}
    q_{old}(S_t,A_t)=(1-\eta) q_{old}(S_t,A_t)+\eta \max_{A_{t+1}} q(S_{t+1},A_{t+1})
\end{equation}
where $\eta$ is the learning rate given to the algorithm as a hyper-parameter.

By repeated interaction over a large number of episodes, the Q-values of all the policy interventions will converge and reveal the expected $MVPF$ at every possible state in the state space and action in the action space. The algorithm below summarizes how the government agent learns.

  \footnotesize
\begin{algorithm}
  \caption{Government Policy Learning Process (Q-learning)}
  \begin{algorithmic}[1]
    \For{\text{each episode $\in$ training episodes}}
     \State \text{Government agent observes the state of the environment after algorithm in Section \ref{timeline} is played}
    \State \text{Government intervenes via exploration or exploitation in an $\epsilon$-greedy strategy}
    \If {Government agent chooses to \textbf{explore}}
    \State \text{Government intervenes by taking a random action from its action space}
    \EndIf
    \If {Government agent chooses to \textbf{exploit}}
    \State \text{Government intervenes by choosing the action that has the highest Q-value}
    \EndIf
    \State \text{Government observes the reward/penalty associated with the action taken.}  
    \State \text{Government updates the Q-value of the chosen action by Equation (\ref{eq:bellmanit}).}
    \EndFor
  \end{algorithmic}
\end{algorithm}
\normalsize
\FloatBarrier
The Q-learning algorithm guarantees convergence to a global optimum if the government interacts with the environment for an infinitely large number of times \citep{watkins1989learning}, \citep{watkins1992q}. For our purposes, we will train the government agent to derive the optimal intervention policy over 1,000,000 episodes.
  \subsection{The Optimal Policy and Policy Insights}
  \label{results}

  After training, the Q-values, i.e. the expected $MVPF$, of each policy intervention under various states of the environment are shown in Table 5. 
  Throughout the subsequent analysis, we will code the states of the environment and government intervention actions as follows:
  \footnotesize
\begin{singlespace}
  \begin{longtable}{
p{1.2cm}
p{15.3 cm}
}
Code&Definition\\ \midrule
\endfirsthead
\endhead
Action 1& Offer state-provided insurance (see Section \ref{intervene} for policy design)\\
Action 2& Ease Solvency Requirements (see Section \ref{intervene} for policy design)\\
Action 3& Provide Awareness Campaigns (see Section \ref{intervene} for policy design)\\
Action 4& Increase Subsidies (see Section \ref{intervene} for policy design)\\
Action 5& Provide Premium Regulations (see Section \ref{intervene} for policy design)\\
Action 6& Offer Disaster Prevention Methods (see Section \ref{intervene} for policy design)\\
Action 7& Increase Reinsurance (see Section \ref{intervene} for policy design)\\
State 1&low awareness,low private supply of cat insurance, affordable premium rates of private insurance\\
State 2&low awareness,high private supply of cat insurance, affordable premium rates of private insurance\\
State 3&low awareness,high private supply of cat insurance, unaffordable premium rates of private insurance\\
State 4&high awareness,low private supply of cat insurance, unaffordable premium rates of private insurance\\
State 5&high awareness,high private supply of cat insurance, unaffordable premium rates of private insurance\\
\midrule
\end{longtable}
 \small
 
\begin{longtable}{
p{1.1 cm}
p{1.8 cm} p{1.5 cm}p{1.5 cm}p{1.5 cm}p{1.5 cm}p{1.5 cm}p{1.5 cm}p{1.5 cm}
}
\caption{Q-values of policy changes under catastrophe risk}\\
 \toprule 
&No Action&Action 1&Action 2&Action 3&Action 4&Action 5&Action 6&Action 7\\ \midrule
\endfirsthead
\caption[]{Q-values (Continued)}\\
 \toprule  
&Action 1&Action 2&Action 3&Action 4&Action 5&Action 6&Action 7&Action 8\\ \midrule
\endhead
State 1&3.709&\cellcolor{black!25}5.657&3.861&\cellcolor{black!10}4.356&0.952&2.999&3.902&3.856\\
State 2&\cellcolor{black!10}0.606&0.156&0.047&\cellcolor{black!25}0.686&0&0.414&0.354&0.296\\
State 3&1.308&1.699&1.123&2.170&\cellcolor{black!10}2.239&\cellcolor{black!25}3.307&2.142&1.132\\
State 4&1.052&1.566&\cellcolor{black!10}1.623&0.962&0.639&0.705&1.054&\cellcolor{black!25}1.788\\
State 5&0.476&0.896&0.179&0.439&0.782&\cellcolor{black!25}2.002&\cellcolor{black!10}1.731&0.083\\
\midrule
\end{longtable}
\end{singlespace}
\normalsize
In dark grey we highlighted the actions which yielded the highest expected $MVPF$ in each state. The second best intervention actions are highlighted in light grey. The government's optimal intervention policy would be to adopt the dark grey shaded action of each state of the environment.

\subsubsection*{State 1}
Under State 1, Table 5 shows that offering state-provided insurance is the optimal intervention. A \$1 spent on increasing state-provided insurance will increase welfare by around 5.66. This result is driven by two factors. First, in a situation where privately supplied catastrophe insurance policies are low, the majority of the individuals in the population are likely to be uninsured. This means that the willingness to pay (WTP) for this policy change would be high as the majority of the population would be willing to pay up to their maximum payable premium $P^{max}_{ti}$. If the government combines this policy change with an increase in catastrophe awareness campaigns, which is the next best policy intervention, individuals' willingness to pay would be even higher as their $P^{max}_{ti}$ would be higher due to higher awareness. Second, increasing state-offered insurance, as compared to most of the other policy interventions, does not impose a sustained financial burden on the government. The government is annually paid a premium rate by insured individuals and, at worst, the government would be obliged to cover the losses of the insured individuals in times of catastrophes. The latter liability is not included in the direct costs of providing state catastrophe insurance.

The next best policy intervention, according to Table 5,  is to provide catastrophe awareness sessions to educate individuals about their true catastrophe risk. A \$1 spent on each would increase welfare by 4.356. As we described above, providing catastrophe awareness sessions increases individuals' awareness of their risk exposure. In a state of the world like that of State 1, where individuals have low awareness, the deviation between individuals' maximum payable premium, $P^{max}_{ti}$, under low awareness with that under full awareness is expected to be very high. Therefore, according to Equation \ref{eq:wtppol3} in Section \ref{wtp}, the willingness to pay by individuals for this policy change is high. Despite that, it is one of the best policy interventions yet not the best intervention for two reasons. First, the cost of administering catastrophe awareness is high in our model and is dependent on the social class of individuals in our society (see Section \ref{cost} for recap). This causes every unit of welfare generated by this policy change to be associated with a higher cost than many other policy changes. The second reason is because the welfare increases are not sustainable. If the government keeps on investing in catastrophe awareness under State 1, the cry wolf effect starts to limit the effectiveness of the policy (see Section \ref{cost} for recap on cry wolf effect). In addition, even if individuals' catastrophe awareness rises significantly, they may still not be able to manage their catastrophe risks as supply of catastrophe insurance is low under this state of the environment, i.e. demand is constrained by supply. 

The worst policy intervention under this state of the environment would be to increase subsidies. According to our model, a \$1 spent on subsidies would increase welfare by 0.952. This is because willingness to pay for this policy change is lower than the cost to the government of administering subsidies. Willingness to pay is low because, under this state, privately supplied catastrophe insurance, when it is available despite its low supply, is affordable. Therefore, willingness to pay for subsidies is not expected to be higher for this policy intervention than other policies. However, the cost to the government of administering this policy is very high. The government would not only have to pay for a part of the premium rates paid by individuals every time they are insured but also bear the consequences of moral hazard. Providing subsidies to individuals gives them the wrong incentives and would further lower their catastrophe awareness. Therefore, the cost to the government of this policy change is higher than individuals' willingness to pay for it.

Another undesirable intervention would be to impose premium regulations on insurers. According to Table 5, a \$1 spent on this policy intervention would increase welfare by 2.99. This is clearly below the welfare increase that would occur to society if the government decided to not intervene at all (3.709 from Table 5). Therefore, a rational government faced with the option of imposing premium regulations and not intervening at all should choose the latter. The reason behind this is that imposing premium regulations are unnecessary under this state as premium rates are affordable and exaggerated premium rates are not the reason why individuals are not buying catastrophe insurance.

\subsubsection*{State 2}
Under State 2, Table 5 shows that the optimal policy intervention is to provide catastrophe awareness sessions. A \$1 spent on providing catastrophe awareness sessions increases welfare by 0.686. The next best action is to take no action at all. A \$1 saved by taking no action instead of intervening with any policy change increases welfare by 0.606.

It is remarkable that the optimal actions under this state both yield less benefit per unit spent making it difficult for the government to make the case for intervention. This observation is due to the fact that, under State 2, conditions are quite favorable and do not invite intervention. Under State 2, while individuals have low catastrophe awareness, there is a high private supply of catastrophe insurance and premium rates are affordable. Under these conditions, individuals can find catastrophe insurance contracts that suit their underestimated understanding of their risk exposure. Therefore, adequate risk management still occurs and the role of government intervention is limited. While investing in catastrophe awareness is the best policy response, conditions pre-intervention were already favorable because not only does every dollar spent increases welfare by less than a dollar but also the welfare generated by the policy intervention is very close to not intervening at all. 

In real catastrophe insurance markets, conditions similar to those of State 2 can occur in times of economic booms without any recent history of catastrophes. People's incomes are high enough for them to purchase catastrophe insurance and firms see an opportunity in the market, which witnessed no recent catastrophe, and therefore increase supply of catastrophe insurance (sometimes irrationally). In light of the status quo, government is satisfied with the situation in the market and is perhaps indifferent between further investing in catastrophe awareness or taking no action at all.

\subsubsection*{State 3}
Under State 3, market prospects are promising for insurers to supply catastrophe insurance yet they are unaffordable to most individuals. Such a scenario may occur in societies with high income inequality where catastrophe insurance may be accessible only to the richest segments of society. Under such high inequality, the majority of the population is likely to be poorly educated and, therefore, have low perception of their true catastrophe protection. 

According to Table 5, the optimal policy intervention is for the government to impose premium regulations on insurers. A \$1 spent on imposing premium regulations would increase welfare by 3.307. Premium regulations include imposing premium ceils on insurers such that maximum premium charged would be actuarially fair. This policy change would increase individuals' purchase of catastrophe insurance as it would help make the offered premium rates be within their affordability i.e. less than or equal to the maximum payable premiums, $P^{max}_{ti}$, of many individuals. If this policy is combined with the provision of catastrophe awareness, the third best policy intervention under this state, premium regulation would be more effective in improving catastrophe protection of society as individuals' $P^{max}_{ti}$ would be high enough to match the efficient level of their true risk exposure.

The second-best policy intervention, according to Table 5, would be to provide subsidies on catastrophe insurance. A \$1 spent on this policy change would increase welfare by 2.24. This result is driven by the low affordability of catastrophe insurance policy despite their high availability; a combination that hints at high income inequality in society. The low-income households, who may represent a sizeable proportion of households in State 3, may not be able to afford catastrophe insurance even if their premiums are actuarially fair. In such a case, the government would subsidize catastrophe insurance to help low-income households obtain catastrophe insurance. A caveat to this policy intervention, which is the reason why it is not the optimal policy intervention under this state, is that it may provide bad incentives and cause moral hazard. To address this caveat, the government agent can combine this policy change with Action 6, increasing the provision of catastrophe prevention technology, which was the fourth-best policy intervention under State 3. Combining these two interventions reduce the effects of potential moral hazard. This policy suggestion is similar to the suggestion proposed by \cite{jscata} who argued that providing flood insurance should also provide poor individuals with options to lower their risk exposure.

The least desirable interventions under State 3 are easing solvency requirements and increasing reinsurance funds. Table 5 shows that these interventions give less welfare per \$1 spent than deciding not to intervene at all. These policy interventions mainly target increasing catastrophe insurance supply which is already high under State 3.

\subsubsection*{State 4}

Under State 4, despite high catastrophe awareness, private supply of catastrophe insurance is not only scarce but, when available, is unaffordable. This indicates that despite a healthy demand condition, a supply-side problem exists in the market. Therefore, it make sense for optimal interventions, according to Table 5, to be to increase reinsurance funds (best intervention) and to ease solvency requirements (second-best intervention). A \$1 spent on increasing reinsurance funds would increase welfare in the society by 1.788. A \$1 spent on easing insurer solvency requirements would increase welfare by 1.623. Increasing reinsurance funds increase insurers' reinsurance accounts which increase insurers' assets as shown in Equation (\ref{eq:assets}) in Section \ref{lossmodel}. An increase in the assets available with insurers can help increase the supply of catastrophe insurance policies. Easing solvency requirements reduces  the amount insurers need to keep in reserves per policy. According to Equation (\ref{eq:supply}) in Section \ref{lossmodel}, supply of catastrophe insurance is inversely related to reserve kept aside per policy. A reduction in the reserve requirement per policy increases the supply of catastrophe insurance. While easing solvency requirements put insurers at a higher risk of insolvency, combining this policy intervention with increasing reinsurance funds helps counter this risk by supplying solvency-constrained insurers with emergency funds when needed.

The third best policy intervention under State 4 is for the government to offer state-provided insurance policies. A \$1 spent on this policy intervention would increase welfare by 1.566. An increase in state-provided catastrophe insurance would increase supply of catastrophe insurance in the society and ameliorates its scarcity under State 4. The fourth best policy intervention, according to Table 5, is to offer disaster prevention technology. A \$1 spent on this policy intervention would increase welfare by 1.054. This result is driven by the fact that disaster prevention technology can help individuals reduce their catastrophe risk exposure and therefore substitute for catastrophe insurance which is scarce under State 4. 

The least effective policy intervention, according to Table 5, is to provide subsidies on catastrophe insurance. A \$1 spent on this policy intervention would increase welfare by 0.639 which is lower than the welfare generated by deciding not to intervene at all. While subsidies address the unaffordability of catastrophe insurance, they do not address the scarcity of insurance policies in the market. Increasing subsidies on catastrophe insurance also come at a high cost to the government: the government is not only obliged to bear the cost of the subsidy itself but also deal with the moral hazard that come with insurance policy subsidization (see Section \ref{cost} for a recap on government policy cost and fiscal externalities).

\subsubsection*{State 5}
Under State 5, while individuals have a high awareness of their risk exposure and demand for catastrophe insurance is in a healthy condition, the supply of catastrophe insurance, despite widely available, is unaffordable to most individuals.

According to Table 5, the optimal intervention is for the government to regulate catastrophe insurance premium rates. A \$1 spent on imposing premium regulations would increase welfare in society by 2.002. Like in State 3, imposing premium regulations ensure that insurers do not charge exaggerated premiums and charge the actuarially fair risk premium rate. This policy intervention would make premium rates more affordable, i.e. within the maximum payable premiums, $P^{max}_{ti}$, of most individuals. As opposed to subsidies, imposing premium caps such that they reflect the true catastrophe risk would not cause moral hazard.

The second best policy intervention, according to Table 5, is for individuals to invest in disaster prevention technology. Given the unaffordability of catastrophe insurance under State 5, investing in disaster prevention technology reduces individuals' catastrophe risk exposure. Namely, it reduces the share of wealth they lose due to catastrophes, $\lambda_{Ri}$. This investment would not only help individuals reduce their risk exposure but also reduce the amount they would need to get in insurance. Therefore, individuals would not need to invest a lot of their income in purchasing catastrophe insurance. 

Under State 5, the most ineffective actions are easing solvency requirements, providing catastrophe awareness sessions, and increasing reinsurance funds. According to Table 5, these policy interventions increase welfare by less per \$1 spent than not intervening at all. Easing solvency requirements and increasing reinsurance funds help increase catastrophe insurance supply, which is already high under State 5. Providing catastrophe awareness sessions do not help in catastrophe preparedness as individuals already have high awareness under State 5.

  \section{Conclusion and Future Work}
  \label{conclusion}
  The goal of this paper was to (1) develop a sequential repeated game of individuals, insurers, and a government that reflects the trends and behavior in real catastrophe insurance markets and (2) explore the potential of RL algorithms, Q-learning in particular, in providing insights as what could be the optimal government intervention policy. The best (worst) interventions, as derived by Q-learning, were those that generated the highest (lowest) welfare per \$1 spent on them. We summarize our findings in the table below:
  \footnotesize
  \begin{longtable}{
p{5.7 cm}|
p{5.5cm}| p{5.5 cm}
}

 \toprule 
\textbf{State of the Environment}&\textbf{Best Interventions}&\textbf{Worst Interventions}\\ \midrule
\endfirsthead
\\
 \toprule  
\textbf{State of the Environment}&\textbf{Best Interventions}&\textbf{Worst Interventions}\\ \midrule
\endhead
\multirow{3}{5.5 cm}{low catastrophe awareness,\\ low supply of catastrophe insurance, \\unaffordable insurance premium rates}&offer state-provided insurance&increase subsidies on insurance\\
    &offer catastrophe awareness campaigns&impose premium regulations\\&&\\
\midrule
\multirow{3}{5.5 cm}{low catastrophe awareness,\\ low supply of catastrophe insurance, \\affordable insurance premium rates}&offer catastrophe awareness campaigns&increase subsidies on insurance\\
    &increase reinsurance&impose premium regulations\\&ease solvency requirements&\\
    \midrule
\multirow{4}{5.5 cm}{low catastrophe awareness,\\ high supply of catastrophe insurance, \\unaffordable insurance premium rates}&impose premium regulations&ease solvency requirements\\
    &increase subsidies on insurance&increase reinsurance funds\\&offer catastrophe awareness campaigns&\\&offer disaster prevention technology&\\
    \midrule
\multirow{4}{5.5 cm}{high catastrophe awareness,\\ low supply of catastrophe insurance, \\unaffordable insurance premium rates}&increase reinsurance funds&increase subsidies on insurance\\
    &ease solvency requirements&impose premium regulations\\&offer state-provided insurance&\\&offer disaster prevention technology&\\
    \midrule
\multirow{3}{5.5 cm}{high catastrophe awareness,\\ high supply of catastrophe insurance, \\unaffordable insurance premium rates}&impose premium regulations&ease solvency requirements\\
    &offer disaster prevention technology&offer catastrophe awareness campaigns\\&&increase reinsurance funds\\
    \midrule
\end{longtable}
\normalsize

Our results contribute to the extensive literature that debate how should the government intervene in catastrophe insurance markets. We do this by reinforcement learning i.e. learning optimal policy intervention by "trial and error". Nascent to economic applications, the use of reinforcement learning, commonly used in Computer Science to train robots and AI-agents in atari games, not only allowed us to derive optimal intervention in a complex non-linear system of individuals and insurers but also evaluate a range of intervention policies by comparing them against each other.

From a policy stand-point, our work, through the creation of a micro-founded artificial society, can provide insights to policymakers wishing to compare alternative intervention policies. Our work closely relates to the recently proposed decision-making framework outlined in \citep{kunreuther2021evaluating} for evaluating intervention strategies in pandemic insurance markets. \citep{kunreuther2021evaluating} suggested producing a library of the possible "events", equivalent to the "states" in our model, and the consequences of each intervention under each event. In light of this, our study serves to set an example on how a set of intervention policies, in any context, can be compared against each other using one underlying theoretical model in an algorithmic manner.

Our study can also be further extended by estimating the distributions of the model's parameters using real data, which was not available at the time this study was conducted. Future work can also further model phenomena like the dependencies between behavioral biases in agents' design. Additionally, the government action space can be extended to include more intervention actions like choice architecture, and the behavioral risk audit proposed by \cite{meyer2017ostrich}.

\setstretch{0.8}
\bibliographystyle{chicago}
\bibliography{ref.bib}
\clearpage
\section{Appendix}
\setstretch{1}
\subsection{Parameter Assignment}
\label{parameters}
\footnotesize
\begin{longtable}{
p{0.35 cm}
p{9.75 cm} p{6cm}
}
\caption{Parameter Assignment}\\
 \toprule 
&Assignment&Justification\\ \midrule
\endfirsthead
\caption[]{Parameter Assignment (Continued)}\\
 \toprule  
&Calibration&Justification\\ \midrule
\endhead
$c_i$&50\% Poor, 30\% Middle, 20\% Upper& to target Gini index of 0.486\\
\midrule
$Y_i$&5,000 if Poor, 12,000 if Middle, 50,000 if Upper&to target a Gini index of 0.486\\
\midrule
$W_{0i}$&uniformly [10,000, 15,000] if Poor, [25,000, 40,000] if Middle, [150,000, 300,000] if Upper& to target Gini index of 0.486\\
\midrule
$a_{0i}$&uniformly [0.001, 0.005] if Poor, [0.003, 0.01] if Middle, [0.005, 0.01] if Upper& poorer individuals can have lower risk perception, due to lower-quality education, than other classes \citep{SMAHSA2020}\\
\midrule
$\lambda_{Ri}$&uniformly [0.6, 1] if Poor, [0.3, 0.6] if Middle [0, 0.3] if Upper&poorer classes can be exposed to higher levels of catastrophe risks \citep{boustan2020effect}\\
\midrule
$\beta_{p}$&uniformly [2,3]&risk perception rises more than twice its initial amount after catastrophe experience \citep{gallagher2014learning}\\
\midrule
$\beta_{oa}$&uniformly [0,1]&to ensure a constant decline in perceived losses with time\\
\midrule
$\beta_{m}$&uniformly [0,1]&no indication in literature about societal distributional assumptions but aiming to reproduce effects of bias in \cite{meyer2017ostrich}\\
\midrule
$\beta_{r}$&uniformly [0,1]&no indication in literature about societal distributional assumptions but aiming to reproduce effects of bias in \cite{meyer2017ostrich})\\
\midrule
$\beta_{s}$&uniformly [0,1]&no indication in literature about societal distributional assumptions but aiming to reproduce effects of bias in \cite{meyer2017ostrich}\\
\midrule
$\beta_{h}$&uniformly [0,1]&no indication in literature about societal distributional assumptions but aiming to reproduce effects of bias in \cite{meyer2017ostrich}\\
\midrule
$\kappa_{0j}$&N(500000,100000)&average initial capital with insurers intended to be 67\% higher than maximum wealth in society\\
\midrule
$\gamma_{0j}$&uniformly [0,1]&to introduce heterogeneity amongst insurers\\
\midrule
$\epsilon_{0j}$&uniformly [0,1]&to introduce heterogeneity to different appetites to adverse market conditions\\
\midrule
$l_j$&uniformly [0,1]&to introduce heterogeneity to different profit margins in the market\\
\midrule
$\rho_j$&uniformly [0.8,1]&to investigate the effects of having different solvency requirements by different insurers\\
\midrule
$\beta'_j$&uniformly [0,1]&no information available in literature on distributional assumptions but we aimed to mimic effects described in \cite{kunreuther2013insurance}\\
\bottomrule
\end{longtable}
\normalsize
\clearpage
\end{document}